\documentclass{egpubl}
\usepackage{pg2026s}

\WsConferencePaper

\usepackage[T1]{fontenc}
\usepackage{dfadobe}  

\usepackage{cite}  
\BibtexOrBiblatex
\electronicVersion
\PrintedOrElectronic
\ifpdf \usepackage[pdftex]{graphicx} \pdfcompresslevel=9
\else \usepackage[dvips]{graphicx} \fi

\usepackage{egweblnk} 

\title[ChronoFuseGS]%
      {ChronoFuseGS: Multi-Temporal Gaussian Fusion with Per-Splat Persistence and Change Visualization}

\author[Batik et al.]
{
\parbox{\textwidth}{\centering 
Tobias Batik\orcid{0000-0002-9764-9409},
Diana Marin\orcid{0000-0002-8812-9719},
Peter Kán\orcid{0000-0001-7437-9955},
Hannes Kaufmann\orcid{0000-0002-0322-9869}
\\[6pt]
TU Wien, Austria
}
}
\usepackage{amsmath}
\usepackage{xcolor}
\usepackage{graphicx}
\usepackage{subcaption}
\usepackage{booktabs}
\usepackage{tabularx}
\usepackage{colortbl}
\usepackage{amssymb}
\usepackage{pifont}
\usepackage{siunitx}
\usepackage[commandnameprefix=always,final]{changes}

\begin{document}

\teaser{
 \includegraphics[width=\linewidth]{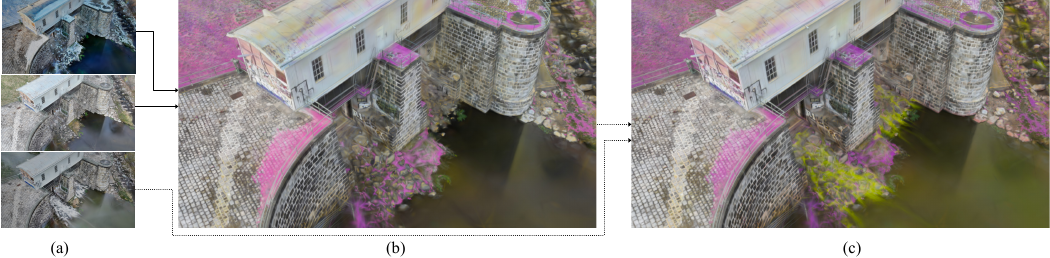}
  \centering
  \caption{ChronoFuseGS incrementally fuses individually trained Gaussian Splatting models and visualizes changes at the Gaussian primitive level. (a) Individual timesteps. (b) Change highlighting between the first two timesteps. (c) After fusion of a third timestep, changes across all three timesteps are highlighted. Highlight colors correspond to timestep presence; persistent parts retain their appearance.}
  
\label{fig:teaser}
}

\maketitle
\begin{abstract}
Reconstructing environments where parts of the scene change between captured image sets poses a challenge for 3D scene reconstruction. We present ChronoFuseGS, a multi-temporal Gaussian Splatting approach that addresses this issue by taking multiple separately trained Gaussian Splatting models, each representing a distinct timestep and partially overlapping in geographic coverage, and merging them into a single combined model. By allowing Gaussians from one timestep to contribute to the reconstruction at other timesteps, our approach leverages data across all captured timesteps to refine persistent parts of the scene. The model supports incremental extension, allowing new timesteps to be added while preserving the existing merged reconstruction. It encodes, for each Gaussian primitive, at which timesteps it contributes to the reconstruction. To support visual exploration of the reconstructed scene, we present a change-aware visualization approach that highlights the parts of the scene that have changed across a user-defined time selection, while preserving the color of persistent parts. Since the persistence encoding operates at the Gaussian primitive level, changes are visualized at sub-object granularity rather than being limited to object-level changes. We evaluate our approach on a real-world outdoor dataset of a flood management area, captured over 7 months across eight recording days and covering seasonal vegetation changes, snow cover, and flooding events, which we make publicly available. Our results demonstrate that the combined model consistently outperforms individually trained single-timestep models in novel-view synthesis quality, recovers structural details absent in the individual reconstructions, and reliably highlights changes in fine details and sub-parts of objects and natural structures.

\ccsdesc[500]{Computing methodologies~Rendering}
\ccsdesc[300]{Computing methodologies~Image-based rendering}
\ccsdesc[300]{Human-centered computing~Scientific visualization}

\printccsdesc   


\end{abstract}

\definecolor{defaultOn}{RGB}{0, 0, 0}
\definecolor{mg_3_1}{RGB}{218, 75, 199}
\definecolor{mg_3_2}{RGB}{218, 142, 74}
\definecolor{mg_3_3}{RGB}{213, 218, 0}
\definecolor{mg_2_1}{RGB}{218, 92, 179}
\definecolor{mg_2_2}{RGB}{218, 182, 13}

\definecolor{s3_4_1}{RGB}{218, 203, 31}
\definecolor{s3_4_2}{RGB}{218, 169, 42}
\definecolor{s3_4_3}{RGB}{218, 137, 38}
\definecolor{s3_4_4}{RGB}{218, 85, 4}

\definecolor{s3_3_1}{RGB}{218, 199, 32}
\definecolor{s3_3_2}{RGB}{218, 159, 81}
\definecolor{s3_3_3}{RGB}{218, 92, 8}

\definecolor{s3_2_1}{RGB}{218, 190, 36}
\definecolor{s3_2_2}{RGB}{218, 109, 23}

\newcommand{\tOff}{{\textcolor{lightgray}{$\circ$ }}}
\newcommand{\tOn}[1][defaultOn]{{\textcolor{#1}{$\bullet$ }}}

\let\added\chadded
\let\deleted\chdeleted
\let\replaced\chreplaced

\definecolor{incrementalcolor}{RGB}{244, 131, 26}   
\definecolor{fromscratchcolor}{RGB}{145, 129, 166}   
\definecolor{mergeshade}{RGB}{150, 150, 150}         
\definecolor{crossshade}{RGB}{100, 100, 100}         
\definecolor{refinementshade}{RGB}{20, 20, 20}    

\newcommand{\colorswatch}[1]{%
    \raisebox{2pt}{\colorbox{#1}{\phantom{\rule{0.05em}{0.05em}}}}%
}

\section{Introduction}
{
Reconstructing large, unknown, and \replaced{evolving}{rapidly changing} outdoor environments is a difficult task that challenges the assumption that the input images represent the scene at a single point in time. This assumption can break down \replaced{for}{in several scenarios. For example,} areas too large to be captured \added{in detail} within a time frame short enough to assume the scene remains unchanged, \replaced{as well as}{or} partially unknown environments, where the required level of detail for different subareas is not known in advance. \replaced{In both cases, repeated recapturing 
is required, and the scene may change between captures}{usually require repeated recapturing}, meaning the input data can no longer be assumed to represent a single unchanged state of the scene, and the resulting reconstruction can no longer be treated as representing a single point in time and rendered to the user as such.
}

{
In particular, we target -- but do not limit ourselves to -- the reconstruction of areas affected by natural disasters such as floods and landslides, where 3D reconstructions provide actionable information for first responders and domain expert teams while the disaster is actively evolving. Taking inspiration from Shneiderman's Visual Information-Seeking Mantra~\cite{shneiderman1996eyes}, we aim for a \added{captur and reconstruction} process in which an overview of the affected area is created first, and detailed reconstructions of subareas are added on demand as new data becomes available. This process may span multiple days as the disaster progresses, altering both natural features and infrastructure. \added{Individual recording periods are chosen such that the scene can be assumed to 
remain static within each period, with changes occurring between captures.}
}

{
Therefore, we propose a multi-time reconstruction approach that builds on Gaussian Splatting~\cite{kerbl2023gaussiansplatting}. Our approach takes multiple individually pre-trained Gaussian Splatting models as input, each trained on a separate image set representing a single timestep, and combines them into a single merged model that is subsequently refined. Gaussians originating from one timestep can thereby contribute to the representation of other timesteps, provided they represent a persistent part of the scene. The model encodes, for each Gaussian, which timesteps it contributes to. A Gaussian is only considered to contribute to a given timestep if it can be validated against at least one image from the corresponding image set. Hence, our model can represent the scene for any user-defined selection of the captured timesteps, not limited to a consecutive sequence. Compared to independently trained single-timestep models, this increases the novel-view synthesis quality, as our results demonstrate in Section~\ref{sec:results}. In addition, our model provides a per-Gaussian encoding of which areas of the scene have changed across timesteps.
}

{
To render novel views for multiple timesteps, we introduce a rendering approach that 
highlights areas that are not present during the selected time period, while retaining the true appearance of persistent parts (Figure~\ref{fig:teaser}). This allows users to explore the whole model without switching between timesteps. At the same time, our change-aware rendering provides a visual indication of when (i.e., during which timesteps) individual parts, objects, and details of the reconstruction are present. Users can therefore quickly identify which areas are persistent across the selected timesteps, which areas may have changed, and at which timestep the change occurred.
}
{
In summary, our contributions are as follows:
\begin{itemize}
    \item A multi-temporal Gaussian Splatting-based reconstruction approach that merges individually pre-trained models into a single, combined model, while encoding per-Gaussian persistence and change across timesteps.
    \item A rendering approach for the multi-temporal model that visually indicates when and where parts of the scene have changed across a user-defined time selection.
    \item A quantitative comparison of our reconstruction against individually trained Gaussian models, and a user study evaluating our visualization approach.
    \item An open-source dataset of a flood management area recorded over 8 days across 7 months, including seasonal vegetation changes, snow cover, and flooding events.
\end{itemize}

\begin{figure*}[ht] 
    \centering
    \includegraphics[width=\textwidth]{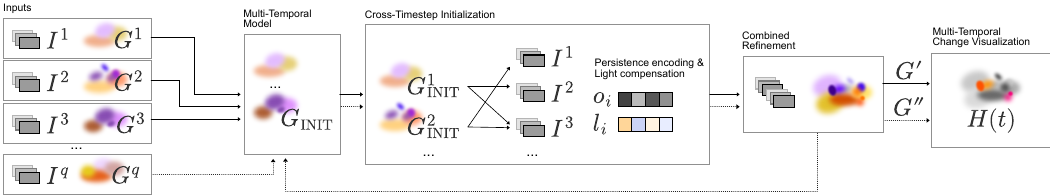}
    \caption{Incremental reconstruction and refinement pipeline. A set of individually trained Gaussian Splatting models and their corresponding image sets are merged into a multi-temporal model $G_\text{INIT}$. This is followed by cross-timestep initialization to estimate the per-Gaussian persistence encoding $o_i$ and light compensation $l_i$, training Gaussians $G^k_\textsc{init}$ originating from $G^k$ against $I^{p \neq k}$, and by a combined refinement step. The refined model enables novel-view synthesis and change highlighting $H(t)$ for any user-defined time selection $t$. Solid arrows indicate the initial computation of $G'$, dashed arrows the incremental fusion of an additional model $G^q$ into $G'$, yielding $G''$.}
   \label{fig:pipeline}
\end{figure*}
\vspace{-6pt}

\section{Related Work}
\subsection{Gaussian Splatting}
Neural radiance fields have become a standard representation for novel-view synthesis, but their implicit formulation often makes training and rendering computationally expensive~\cite{mildenhall2020nerf}. 3D Gaussian Splatting (3DGS) introduced an explicit radiance-field representation based on anisotropic 3D Gaussians optimized directly from posed images~\cite{kerbl2023gaussiansplatting}. Each Gaussian stores position, covariance, opacity, and view-dependent appearance, and the scene is rendered using a visibility-aware differentiable rasterizer. This explicit formulation combines high-quality view synthesis with real-time rendering, making 3DGS attractive for interactive graphics applications and for downstream tasks that require direct access to scene primitives.

\subsection{Multi-Timestep and Dynamic Reconstruction}
A large body of recent work extends 3DGS from static scenes to dynamic reconstruction. One common strategy is to represent a scene in a canonical space and learn a time-dependent deformation field that maps the canonical Gaussians to each observed frame~\cite{yang2024deformable3dgs}. 
A second line of work represents time more directly by introducing 4D or time-aware Gaussian primitives~\cite{wu20244dgaussians}. These approaches can render a scene at different timestamps by querying a shared spatio-temporal representation. 
Both families are well-suited to continuous temporal interpolation of a single coherent dynamic capture, but are not designed for our setting of repeated, independently captured reconstructions of approximately the same environment, where each timestep should remain an identifiable reconstruction while benefiting from shared cross-timestep information.

Long-term mapping and scene-update methods are more closely aligned with our setting. Instead of reconstructing every timestamp independently, these methods reuse an existing Gaussian map, detect stale or changed regions, and perform targeted updates~\cite{cheng2025ltgs,fu2025gslts,yugay2025game}. These approaches show that temporal reuse is valuable for efficiency and consistency, but most existing systems maintain a single evolving map rather than a structured collection of timestep-specific reconstructions.

The closest \added{method} to our reconstruction approach is ChronoGS~\cite{wang2026chronogs}, which \added{similarly} represents multiple timesteps within a unified \replaced{model and shares information across timesteps to refine the persistent regions. ChronoGS jointly optimizes a single anchor scaffold across all timesteps, using an MLP to decode features into Gaussian attributes.}{scaffold by modeling cross-timestep variation through a learned MLP. Unlike our approach, ChronoGS requires joint optimization from scratch across all timesteps simultaneously.} In contrast, our approach takes individually pre-trained Gaussian models as input, allowing \replaced{each model to be}{the computational load to be distributed across independent models that can be} trained, pruned, and optimized in parallel before merging. Furthermore, new timesteps can be incorporated incrementally without rebuilding the complete multi-temporal model from scratch. 
\replaced{Moreover, while ChronoGS uses MLP-predicted activation to render individual timesteps, our explicit per-Gaussian persistence encoding supports rendering arbitrary timestep selections and directly highlighting when and where changes occur.}{Additionally, our approach supports multi-timestep rendering with explicit per-Gaussian persistence encoding, indicating which areas can be validated for each timestep and highlighting which parts have changed -- functionality not addressed by ChronoGS.}

\subsection{Change Detection in Gaussian Splatting}

Change detection with 3DGS has recently emerged as a natural consequence of explicit radiance-field reconstruction. A common strategy is render-then-compare: a pre-change 3DGS model is rendered from post-change camera poses and compared against newly captured images. 3DGS-CD follows this approach by localizing post-change images against a pre-change Gaussian model, extracting image features from rendered and observed views, and fusing view-wise differences into object-level changes~\cite{lu20253dgs}. MV3DCD similarly targets multi-view change detection, but learns an additional change channel in the Gaussian representation so that change masks can be rendered from unseen viewpoints~\cite{galappaththige2025mv3dcd}. 
As an alternative to render-space comparisons, GS-Diff compares Gaussian reconstructions directly in primitive space~\cite{galappaththige2026gsdiff}. By accounting for uncertainty and visibility, primitive-space differencing reduces false positives caused by poorly observed regions or reconstruction drift. However, GS-Diff is evaluated on controlled scenes with clearly defined object-level changes, and it is unclear how primitive-space comparison performs on large outdoor environments where complex natural structures are approximated by a sparse number of Gaussians at low reconstruction detail.

While both render-then-compare and primitive-space methods derive a change signal that is mapped back onto Gaussian primitives as a post-hoc label, neither approach uses the cross-timestep image comparison as a training signal that modifies the reconstruction itself. In contrast, our method propagates gradients from cross-timestep image-space comparisons back through the differentiable renderer to jointly update geometry and per-Gaussian persistence parameters, making change encoding an inherent property of the reconstruction rather than a derived label.

Our work is positioned at the intersection of these directions. Dynamic 3DGS methods provide temporal regularization and shared structure; long-term mapping methods provide mechanisms for reusing and updating Gaussian maps; and change-detection methods provide render-space and primitive-space comparison strategies. 
We focus on fusing multiple individually trained Gaussian models -- each representing a distinct timestep -- into a single combined model, where cross-timestep information is used to improve reconstruction quality and enable fine-grained visualization of scene changes at the Gaussian level.

\begin{figure*}[t] 
    \centering
    \begin{subfigure}[b]{0.33\textwidth}
        \centering
        \includegraphics[width=\textwidth]{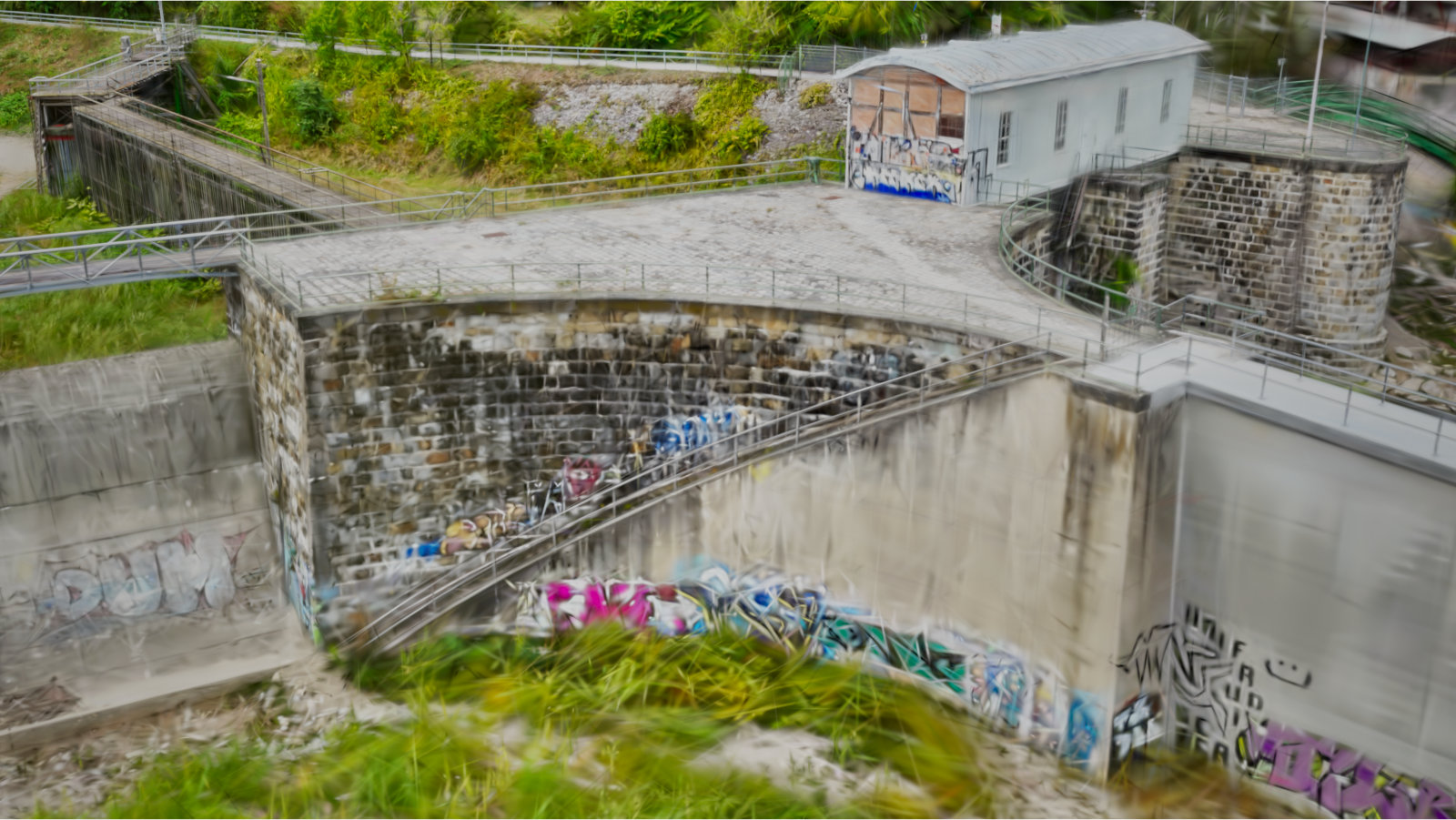}
        \caption{$[$\tOn \tOff \tOff \tOff$]$}
        \label{fig:total_combined_figure_1}
    \end{subfigure}
    \begin{subfigure}[b]{0.33\textwidth}
        \centering
        \includegraphics[width=\textwidth]{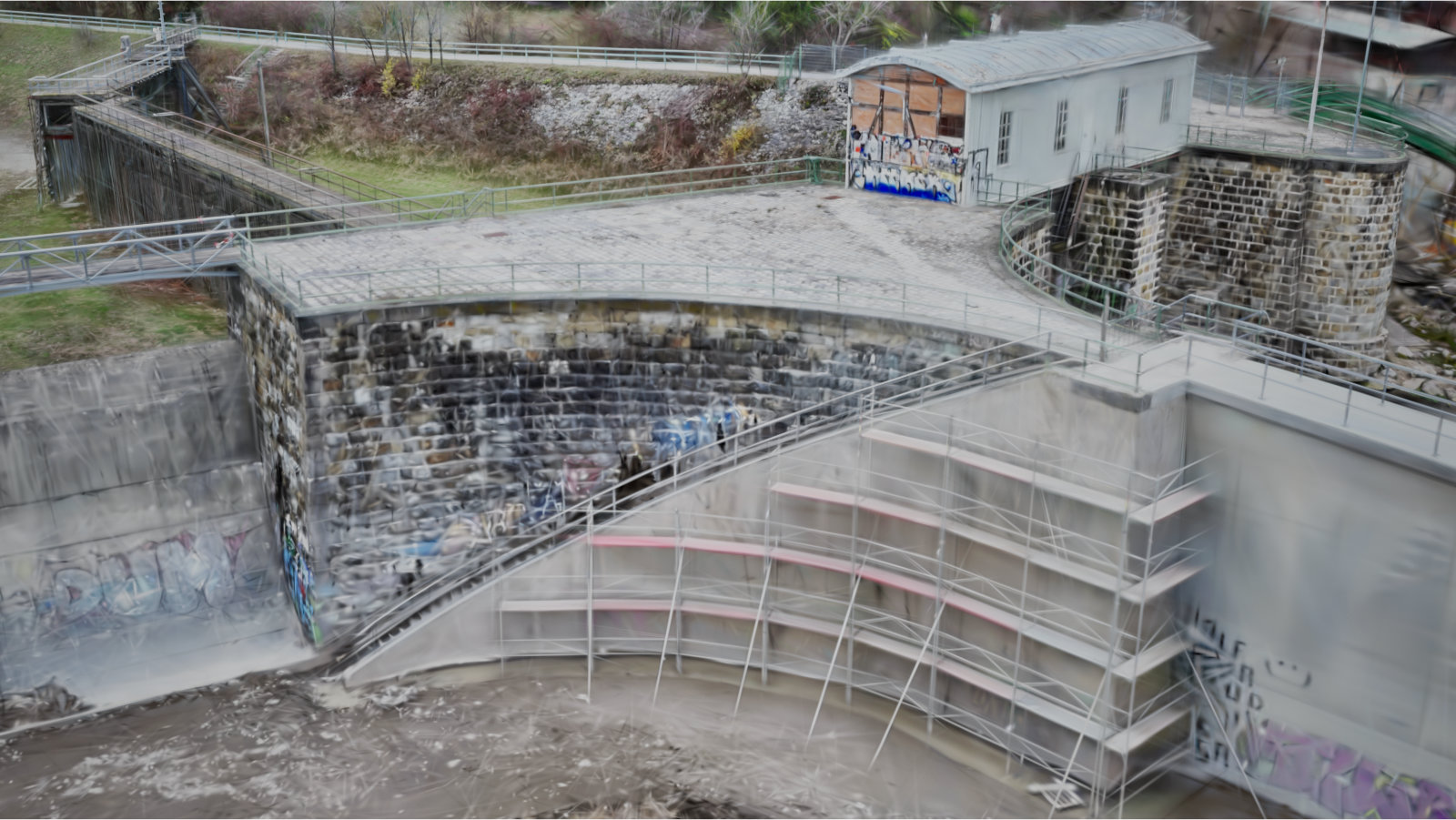}
        \caption{$[$\tOff \tOff \tOn \tOff$]$}
         \label{fig:total_combined_figure_2}
    \end{subfigure}
    \begin{subfigure}[b]{0.33\textwidth}
        \centering
        \includegraphics[width=\textwidth]{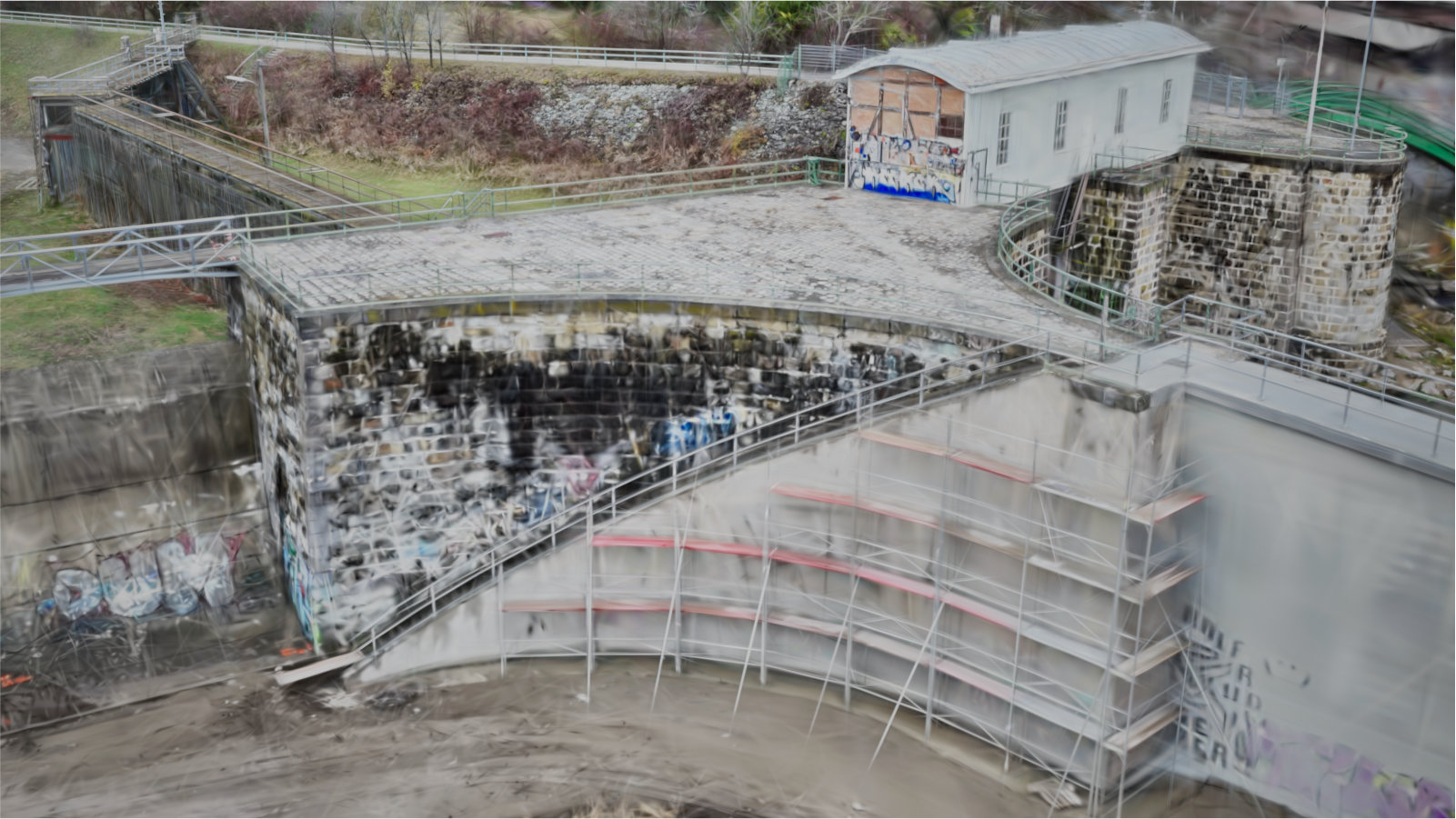}
         \caption{$[$\tOff \tOff \tOff \tOn$]$}
          \label{fig:total_combined_figure_3}
    \end{subfigure}
    \begin{subfigure}[b]{0.33\textwidth}
        \centering
        \includegraphics[width=\textwidth]{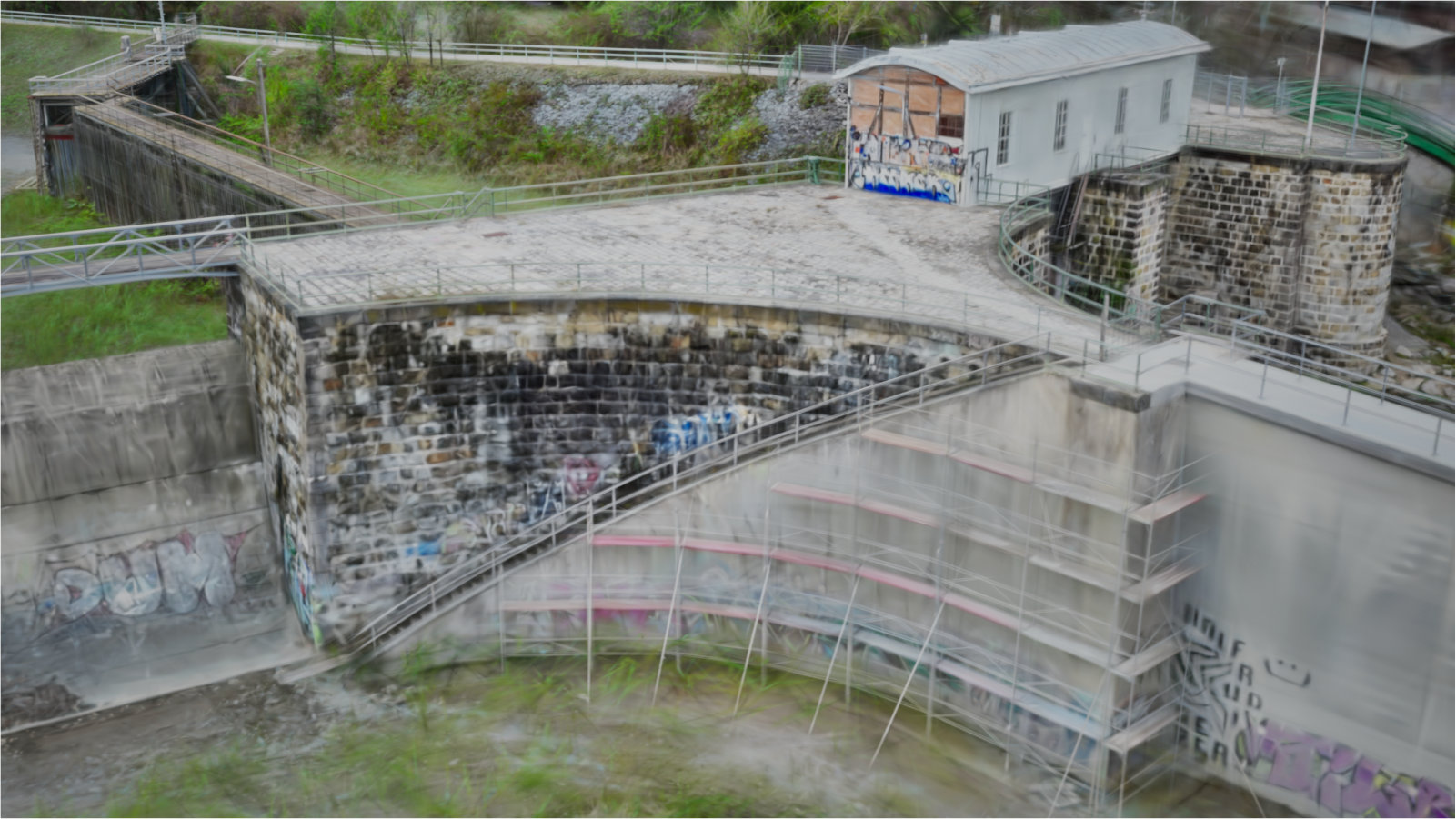}
         \caption{$[$\tOn \tOff \tOn \tOn$]$}
         \label{fig:total_combined_blended}
    \end{subfigure}    
    \begin{subfigure}[b]{0.33\textwidth}
        \centering
        \includegraphics[width=\textwidth]{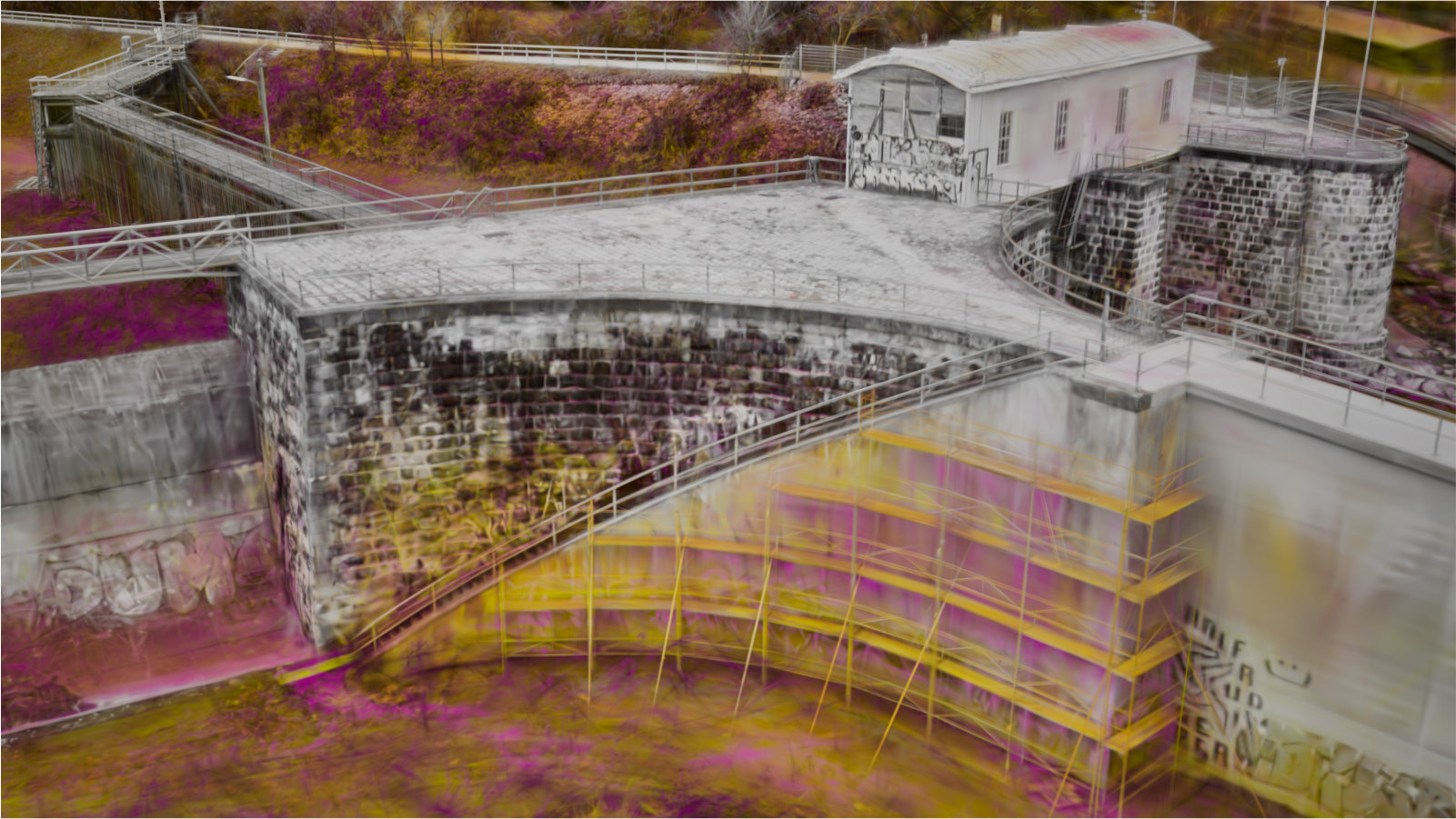}
         \caption{$[$\tOn[mg_3_1] \tOff \tOn[mg_3_2] \tOn[mg_3_3]$]$}
    \end{subfigure}
    \begin{subfigure}[b]{0.33\textwidth}
        \centering
        \includegraphics[width=\textwidth]{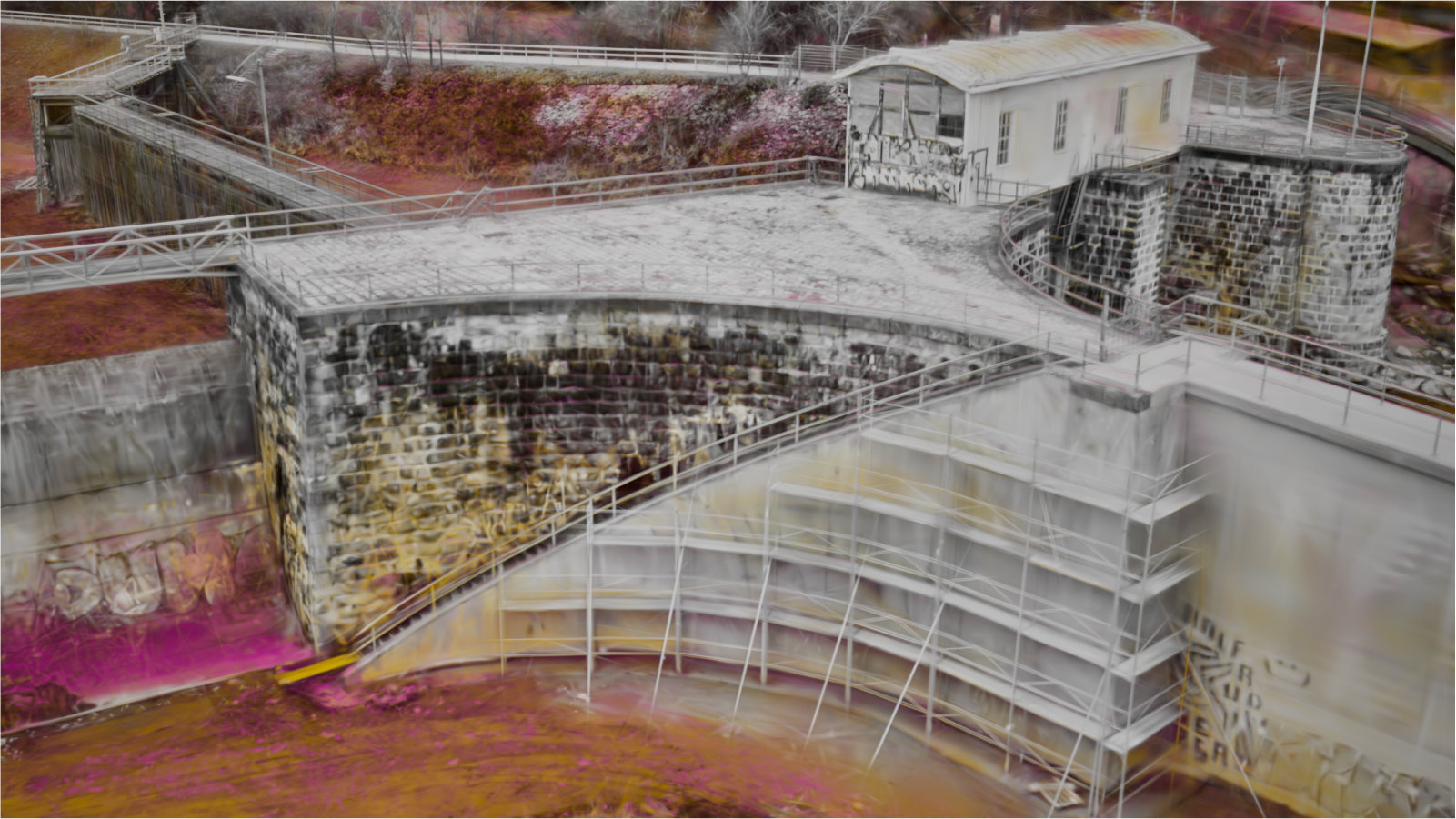}
         \caption{$[$\tOff \tOff \tOn[mg_2_1] \tOn[mg_2_2]$]$}
    \end{subfigure}
\caption{Individual timestep renderings and change highlighting for different time selections. (a)--(c) show the multi-temporal model rendered at $t_1$, $t_3$, and $t_4$; (d) shows the blended multi-time view ($t_1$, $t_3$, and $t_4$); (e)--(f) show change highlights with Gaussian color $c_i$ converted to grayscale. In (e), the scaffolding is absent at $t_1$, so it is rendered in the combined highlight color of $t_3$ and $t_4$. In (f), the time selection is restricted to the last two timesteps, and the scaffolding appears in grayscale as it changed minimally between them. \tOn indicates enabled timesteps, with color corresponding to timestep presence in (e)--(f); \tOff indicates a disabled timestep.
}
    \label{fig:total_combined_figure}
\end{figure*}
\vspace{-6pt}

\section{Multi-Temporal Gaussian Fusion}
{
Our approach takes multiple Gaussian Splatting models $G^1, G^2, \dots$, each representing a distinct point in time $t_1, t_2, \dots$ and individually trained on a corresponding set of ground-truth images $I^1, I^2, \dots$. These models are combined into a single refined model $G'$ that supports novel-view synthesis for each timestep. New timesteps can be incorporated incrementally by fusing an additional pre-trained model $G^k$ into the existing model $G'$, yielding an updated model $G''$. Our merging and cross-timestep fusion process consists of three steps, outlined in the following subsections. First, we bring all Gaussians into a shared coordinate space and combine them into a single multi-temporal Gaussian model. 
Second, in a cross-timestep initialization, we estimate light compensation and initialize the per-Gaussian persistence between all pairs of timesteps.
Third, we perform a combined optimization run to refine the multi-temporal Gaussian model. The pipeline is illustrated in Figure~\ref{fig:pipeline}.
}

\subsection{Multi-Temporal Model}
{  
Since the merged model $G'$ represents multiple timesteps, we store, for each Gaussian $g_i$, its contribution to the reconstruction at each timestep via the opacity manipulation vector $o_i$. The length of this vector equals the number of timesteps in $G'$. A value of $o_i^p \approx 1$ indicates that the Gaussian is visible at time $t_p$, while $o_i^q \approx 0$ results in an invisible Gaussian at $t_q$.
To compensate for light differences between individual timesteps, similar to Kulhanek et al.~\cite{kulhanek2024wildgaussians}, we store, for each Gaussian $g_i$, the light compensation matrix $l_i$, composed of an RGB vector for each timestep of $G'$. Hence, $l_i$ is of shape $3 \times |t|$. Since the direction of light and, therefore, various elements, e.g., shadows, can change between individual timesteps, a single RGB for all Gaussians originating from the same timestep is not enough. Note that for $o_i^p$ and $l_i^p$ we use the subscript $i$ to denote the correspondence to the $i^{th}$ Gaussian and the superscript $p$ to denote the timestep $t_p$. Besides the time-dependent parameters $o$ and $l$, our Gaussian primitives store the same parameters as the original 3DGS: position, opacity $\alpha$, rotation, scale, and color $c$.
}

{
To combine our initial reconstructions, we first have to bring them into a shared coordinate space. \deleted{Formally, we find the affine transformation to transform $G^p$ into the combined reconstruction $G'$.} In the absence of precise real-world positioning data for the training images, we compute another Structure from Motion~(SfM) matching step to register a subset of our new ground-truth images to the corresponding SfM model of $G'$. 
Next, we transform all splats of $G^p$ to fit the coordinate space of $G'$ and merge them into the multi-temporal Gaussian set $ G'_{\textsc{init}}$. Each Gaussian's $o_i$ vector is initialized with a value of 1 for the position corresponding to the timesteps it was initially trained on; all other elements of $o_i$ are initialized with 0. The light compensation, $l_i$, is initialized with 1 for all components, causing no initial effect. 
}

{ 
Rendering our multi-temporal Gaussian model follows the original 3DGS approach~\cite{kerbl2023gaussiansplatting}. We project Gaussians into 2D and blend the $N$ depth-ordered Gaussians that overlap a pixel. Each Gaussian contributes to the pixel color with its color, $c_i$, weighted by its opacity $\alpha_i$ and the accumulated transmittance of Gaussians in front of it. We extend this formulation by incorporating $o_i^p$ and $l_i^p$, yielding the timestep-dependent pixel color $C(p)$ as:
\begin{align}\label{equ:singleTimeRendering}
    C(p) &= \sum_{i \in N} l_i^p \odot c_i \alpha_i o_i^p
        \prod^{i-1}_{j=1}(1-\alpha_j o_{j}^p)
\end{align}
with $\odot$ representing the Hadamard product. For a detailed explanation of projecting Gaussians onto the screen plane, we refer to the original 3DGS publication.
}

\subsection{Cross-Timestep Initialization}\label{ssec:met_init}
{
The cross-timestep initialization process computes initial values for the time-dependent parameters $o$ and $l$ by optimizing each subset of Gaussians originating from the same timestep against the ground-truth images of all other timesteps, while disabling the remaining Gaussians of $G'$. \added{During this stage, $\alpha$ remains fixed at the values established during the individual per-timestep training.} More precisely, for every subset of Gaussians originating from $t_p$, we optimize $o_i^{q}$ and $l_i^{q}$ for all $q \neq p$ against $I^q$. Gaussians representing persistent parts of the scene should result in $o_i^q \approx 1$ with a light compensation $l_i^q$ such that $C(q) \approx I^q_k$, where $k$ is the index of the individual training images of $I^q$. Gaussians representing changed or absent parts should result in $o_i^q \approx 0$. To guide this optimization, we use the spatial feature difference map from LPIPS~\cite{zhang2018perceptual} as a per-pixel similarity weight $\epsilon$, with $\epsilon \approx 0$ indicating high similarity and $\epsilon \approx 1$ indicating no similarity. The following loss is evaluated per pixel and minimized over all training images $I^q_k$:
\begin{align}\label{equ:init_los}
    \mathcal{L}_{\textsc{init}} = 
        (1- \epsilon)\lvert C(q) - I^q_k \rvert + 
        \epsilon \lvert b - C(q) \rvert \, \text{,}
\end{align}
where $b$ is a randomly selected background color, chosen independently for each optimization step to decouple $l$ and $o$. Otherwise, Gaussians representing changed areas may adapt their color to the background instead of converging to $o_j^q \approx 0$. We initialize $o_i^{q \neq p} = \delta o_i^p$ with $\delta < 1.0$ to prevent Gaussians close to the camera from fully occluding those further back. In the first iterations, we set $\epsilon = 0.5$, and uniformly and gradually transition to the LPIPS-based per-pixel similarity as the optimization progresses.
To prevent $l$ from encoding arbitrary color changes, we constrain its components to cold/warm and luminance shifts, limiting the degrees of freedom of the per-Gaussian light compensation.
}

\begin{figure}[t]
    \centering
    \begin{subfigure}[b]{0.49\columnwidth}
        \centering
        \includegraphics[width=\textwidth]{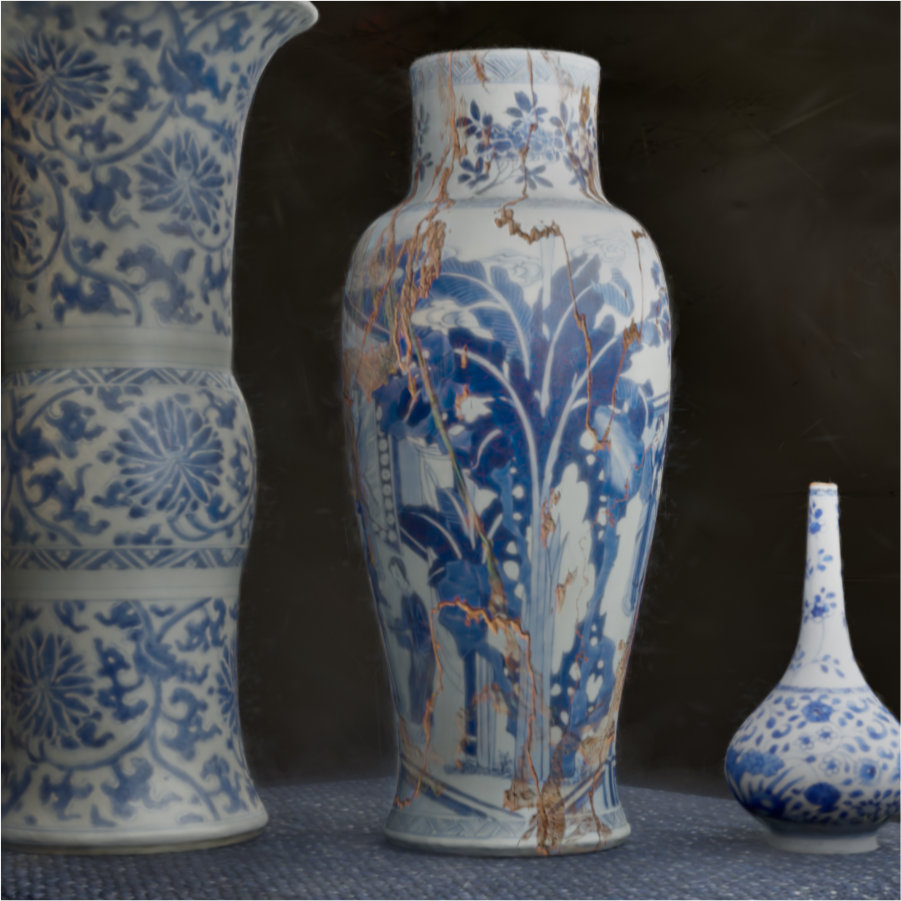}
        \caption{$[$\tOn \tOn \tOn \tOn$]$}
        \label{fig:vase_all_timesteps_blended}
    \end{subfigure}
    \begin{subfigure}[b]{0.49\columnwidth}
        \centering
        \includegraphics[width=\textwidth]{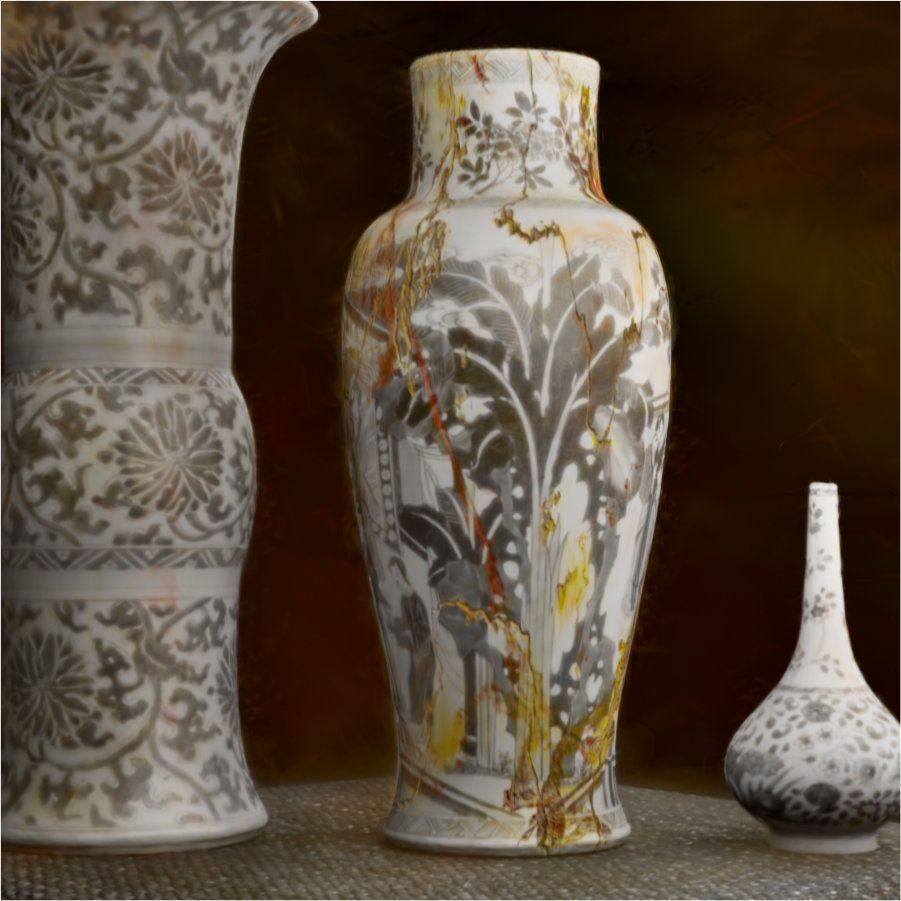}
        \caption{$[$\tOn[s3_4_1] \tOn[s3_4_2] \tOn[s3_4_3] \tOn[s3_4_4]$]$}
    \end{subfigure}
     \caption{Reconstruction of the Baluster Vase dataset across four timesteps, with new cracks progressively added to the central vase. (a) Blended multi-time view with all timesteps selected; (b) the same view with change highlighting, where persistent parts are rendered in grayscale and crack colors indicate at which timesteps they are present. \tOn indicates an enabled timestep, with color corresponding to timestep presence in (b).}
    \label{fig:vases_all_times}
\end{figure}

\subsection{Combined Refinement}
{
After the cross-timestep initialization, Gaussians from different timesteps may overlap in the same region. Since appearance is achieved by alpha-blending, the combined optimization step refines the Gaussian parameters to find the linear combination of Gaussians that best approximates the scene at each timestep.
For the refinement, we run a combined optimization close to a classic 3DGS optimization, training $G'$ against all training images across all timesteps.  \added{Since scene changes between timesteps are encoded by the 
differences between corresponding elements of $o_i$ determined during cross-timestep initialization, we aim to preserve this encoding during refinement.} We \added{therefore} extend the 3DGS loss by a penalty term $\mathcal{L}_\textsc{change}$ that constrains the sequential differences between elements of $o_i$ relative to \replaced{their values $\tilde{o}_i$ defined after the cross-timestep initialization}{their values at the start of the optimization $\tilde{o}_i$}. The sequential difference is defined as $d(o_i) = [o_i^p - o_i^{p+1}, \dots]$. Hence, the refinement loss is:
\begin{gather}
    \mathcal{L}_\textsc{refine} = (1 - \lambda)\mathcal{L}_1 + 
    \lambda \mathcal{L}_\textsc{d-ssim} + 
    \sigma \mathcal{L}_\textsc{change}
\end{gather}
where $\mathcal{L}_{\textsc{change}} = \text{MAE}(d(\tilde{o}), d(o))$, with $\lambda = 0.2$, and $\sigma = 0.1$, \added{weighting the image similarity and regularization terms respectively}. Since Gaussians in $G'$ can overlap because they originate from different initial Gaussian sets, we multiply $\alpha$ by $0.5$ at the start of the optimization, encouraging the model to primarily resolve overlaps through $\alpha$ while preserving the persistence encoding in $o$. 
\added{Additionally, as training images are sampled randomly from all timesteps, $\alpha_i$ cannot be adjusted to compensate for $o_i$ at a specific timestep without simultaneously affecting all others, keeping the two parameters functionally disentangled.}
\replaced{Results are}{Novel-view synthesis using our multi-temporal model for individual timesteps is} shown in Figures~\ref{fig:total_combined_figure_1}--\ref{fig:total_combined_figure_3}.
}

\begin{figure}[t]
    \centering
    \begin{subfigure}[b]{0.495\columnwidth}
        \centering
        \includegraphics[width=\textwidth]{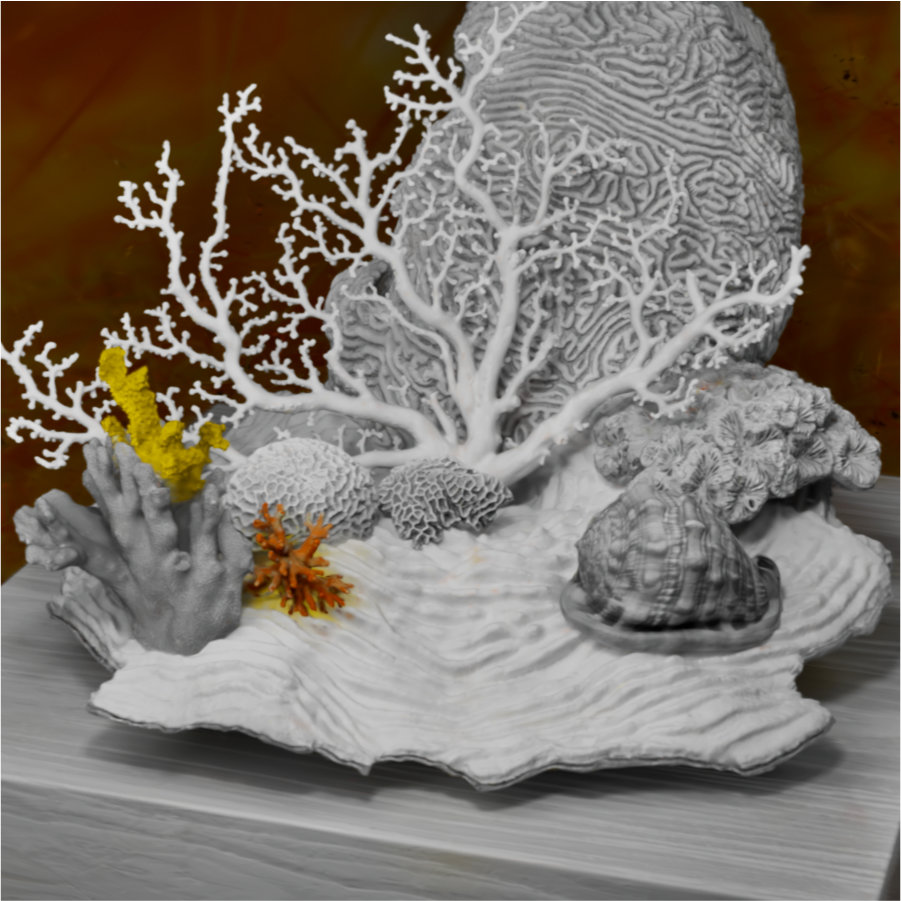}
        \caption{$[$\tOff \tOff \tOn[s3_2_1] \tOn[s3_2_2]$]$}
    \end{subfigure}
    \hfill
    \begin{subfigure}[b]{0.495\columnwidth}
        \centering
        \includegraphics[width=\textwidth]{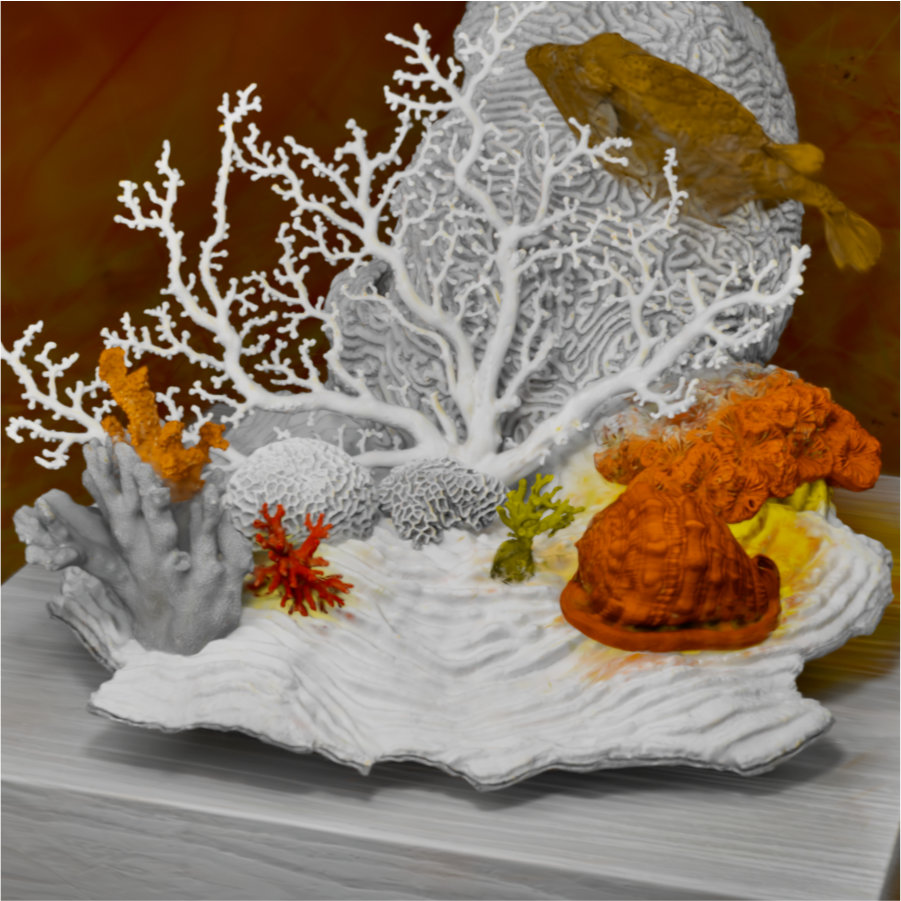}
        \caption{$[$\tOn[s3_4_1] \tOn[s3_4_2] \tOn[s3_4_3] \tOn[s3_4_4]$]$}
    \end{subfigure}
    \caption{Reconstruction of the Coral data with change visualization and different time selections. The Gaussian color $c_i$ is converted to grayscale. In (a), the last two timesteps are enabled, and in (b), changes across all four timesteps are highlighted.
    \tOn indicates an enabled timestep, with color corresponding to timestep presence; \tOff indicates a disabled timestep.
    }
    \label{fig:result_synthetic}
\end{figure}

\begin{figure*}[t]
    \centering
    \begin{subfigure}[b]{0.33\textwidth}
        \centering
        \includegraphics[width=\textwidth]{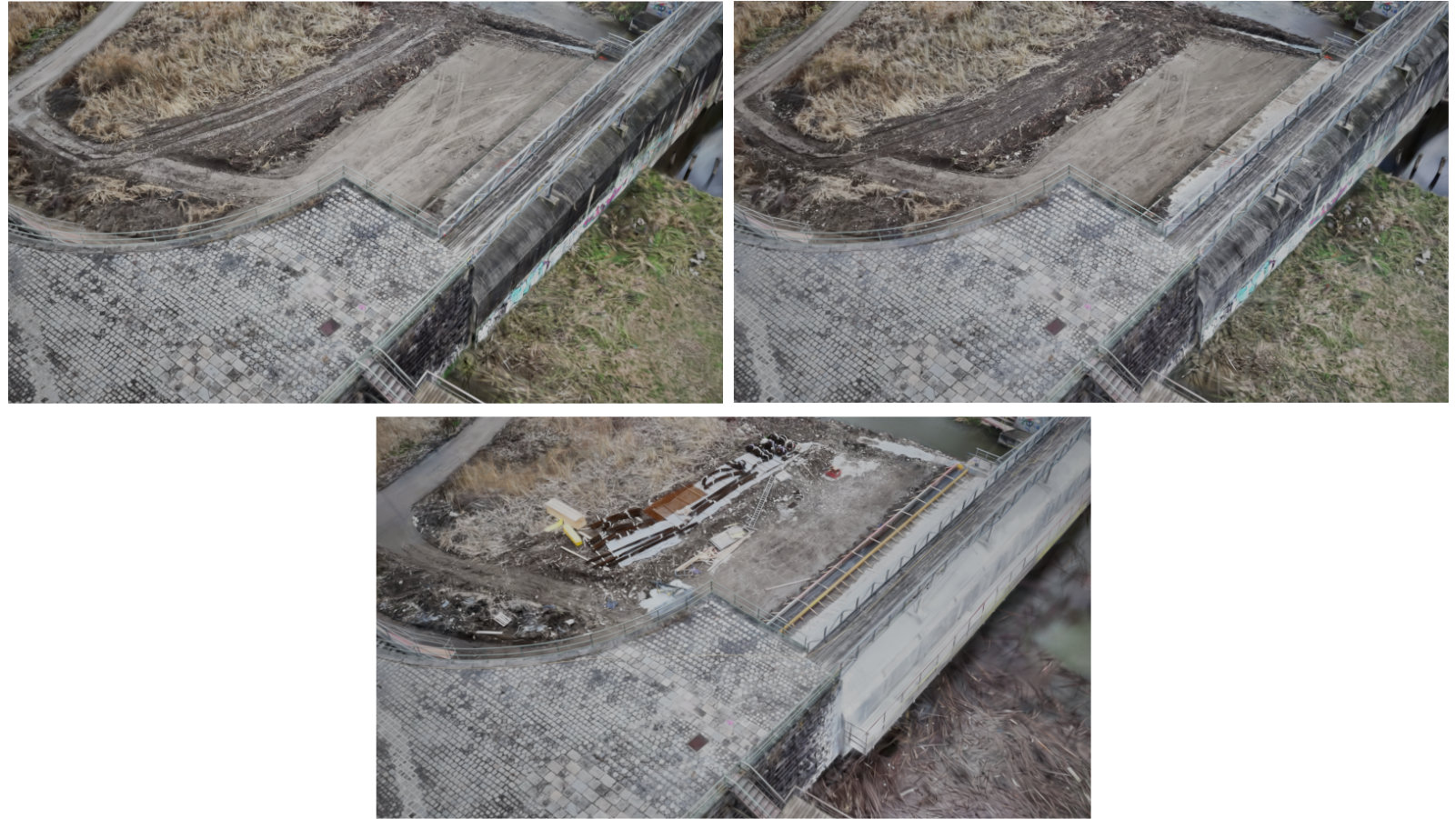}
        \caption{}
        \label{fig:time_selection_highlight_more}
    \end{subfigure}
    \hfill
    \begin{subfigure}[b]{0.33\textwidth}
        \centering
        \includegraphics[width=\textwidth]{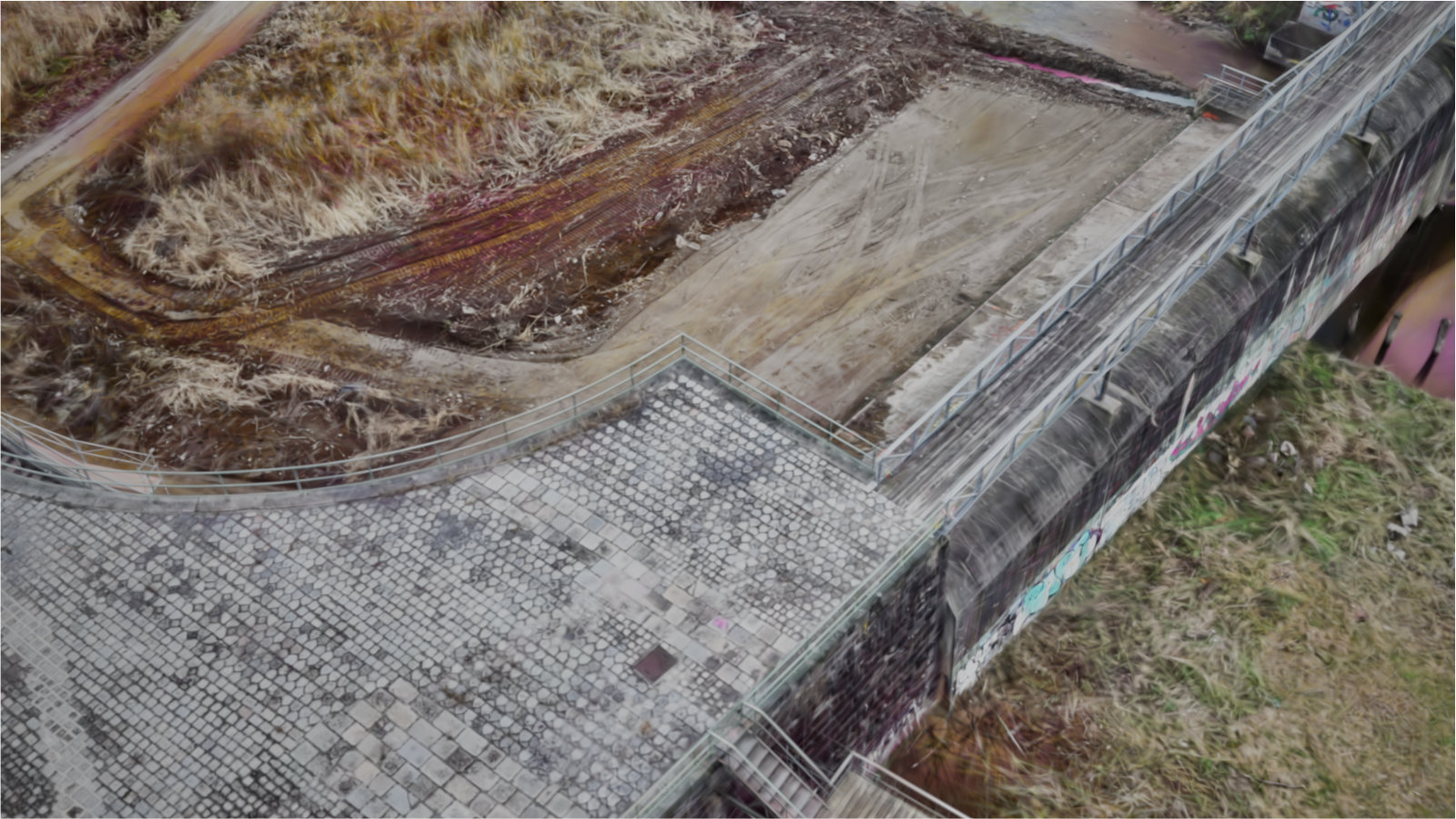}
        \caption{$[$\tOn[mg_2_1] \tOn[mg_2_2] \tOff$]$}
        \label{fig:time_selection_highlight_little}
    \end{subfigure}
        \hfill
    \begin{subfigure}[b]{0.33\textwidth}
        \centering
        \includegraphics[width=\textwidth]{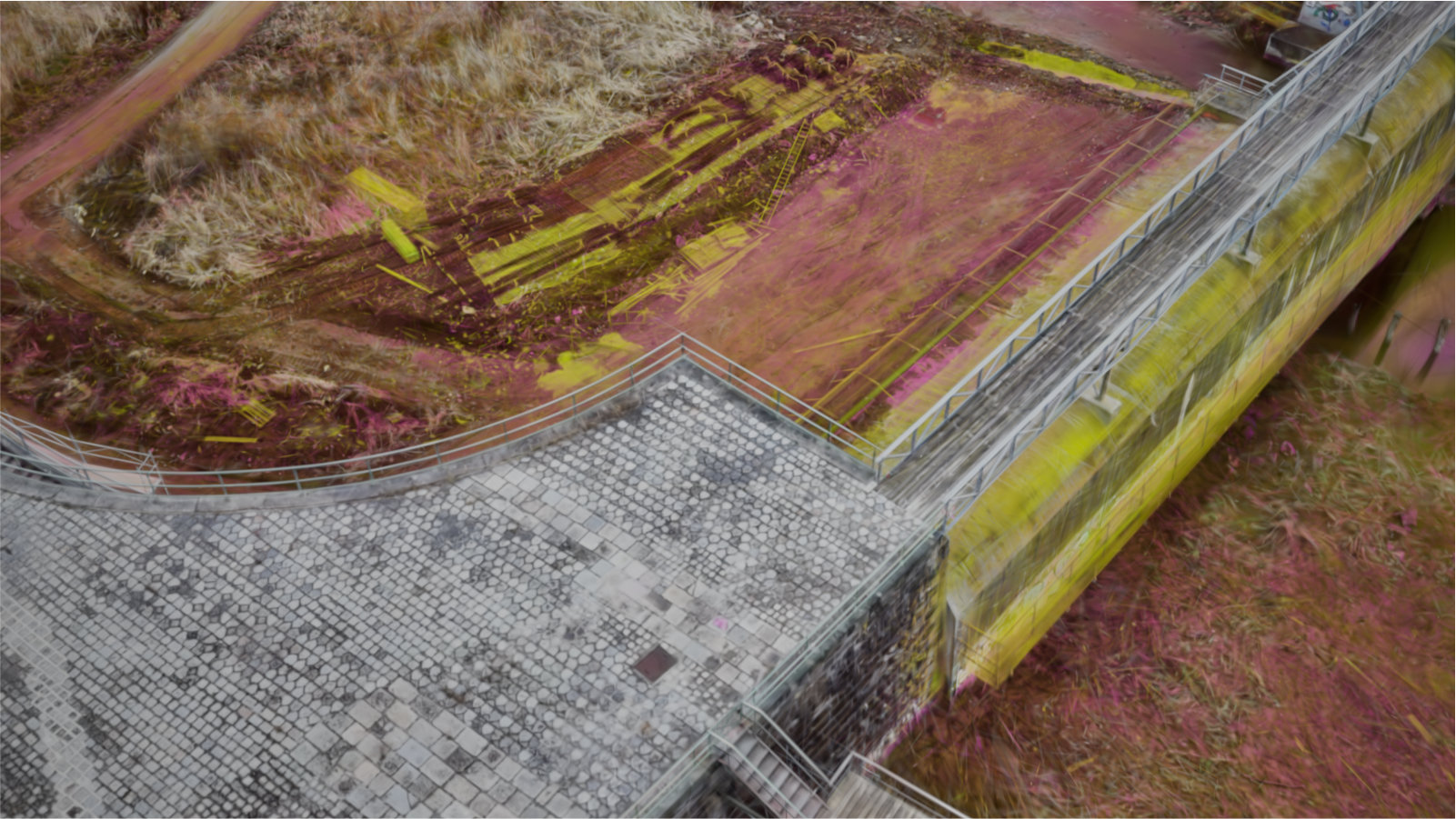}
        \caption{$[$\tOn[mg_3_1] \tOn[mg_3_2] \tOn[mg_3_3]$]$}
        \label{fig:time_selection_highlight_all}
    \end{subfigure}
    
    \caption{Individual timestep renderings and change highlighting for different time selections. (a) Individual timestep renderings at $t_1$, $t_2$, and $t_3$. (b) $t=[1,1,0]$, with little change between the first two timesteps, resulting in a nearly unaltered representation. (c) $t=[1,1,1]$, extending the selection to all three timesteps, highlighting changes across all timesteps.  \tOn indicates an enabled timestep, with color corresponding to timestep presence; \tOff indicates a disabled timestep.}
    \label{fig:time_selection_highlight}
\end{figure*}
\vspace{-6pt}

\section{Multi-Temporal Change Visualization}\label{ssec:rendering}
{
Our trained multi-temporal Gaussian Splatting model represents the reconstructed area for multiple timesteps, supporting partial and varying spatial coverage across timesteps.
To efficiently represent the Gaussian model for user exploration, we outline an approach to render user-defined time ranges. Additionally, in our rendering approach, non-persistent areas between any selected timesteps can be highlighted on the Gaussian level.
}

{
To render the multi-temporal Gaussian model for a user-defined time range, we encode the time selections in a vector $t = [t_0, t_1, \cdots, t_n]$, where components equal to 1 signify that the respective timestep should be visible, and components equal to 0 signify that the timestep is not visible. We could allow every value between 0 and 1, but we found in our testing that values between 0 and 1 were hard to interpret by users. Hence, we use a normalized time selection vector, written as $\hat{t} = t / \sum_{t_i \in t} t_i$.
For a true color representation of our scene for a time selection $t$, we extend Equation~\ref{equ:singleTimeRendering} to 
\begin{align}\label{equ:multiTimeColor}
    C(\hat{t}) &= \sum_{i \in N} B_i(\hat{t}) 
        \prod^{i-1}_{j=1}(1-\alpha_j  o_j \cdot \hat{t}) \\
    \text{with} \quad
    B_i(\hat{t}) &= (l_i \cdot \hat{t}) \odot c_i \alpha_i (o_i \cdot \hat{t})\text{.}
\end{align}
Here $\cdot$ represents the dot product and $\odot$ the Hadamard product. 
$C(\hat{t})$ results in a weighted average of the appearance of the splats over the time selection $t$.
Objects of the scene that are removed in one of the selected timesteps would appear semi-transparent, while parts of the captured scene that stay constant would render as solid.
Alternatively, to avoid semi-transparent parts for objects that are non-persistent between timesteps, we can use $\max(o_i \odot t)$ instead of $o_i \cdot \hat{t}$. 
The visualizations computed with Equation~\ref{equ:multiTimeColor} provide a true color overview of the scene, but changes in small details or changes in complex scenes can be hard to spot for the user, as shown in Figures~\ref{fig:total_combined_blended} and~\ref{fig:vase_all_timesteps_blended}.
}

\subsection{Change Visualization}\label{ssec:change_highlight}
We aim to provide the observer with visual feedback on which subareas of the reconstruction are confirmed by the input data and which subareas may have changed, while preserving the overall appearance of the captured scene. For areas that have changed during the selected timespan, we alter the color to indicate that something changed and at which timesteps it happened. 
However, input images may not cover all subareas at every timestep, and objects may be occluded at some timesteps. Such areas are treated as non-validatable, meaning that their Gaussians may be absent ($o_i^p \approx 0$) for some of the
timesteps. This means non-persistence is not a single event and cannot be encoded by a single point along a linear color gradient.
A color lookup table of all potential combinations of appearances is equally infeasible: four timesteps would already require 14 colors, exceeding the recommended limit for categorical color scales~\cite{ware2019information}, and since $o_i^p$ is a continuous value rather than a binary presence flag, a lookup table would require a user-defined threshold to classify each Gaussian as present or absent. Therefore, we visualize non-persistence in a color space where colors corresponding to timesteps at which a Gaussian is absent are filtered out, leaving fully persistent Gaussians visually unaltered.

{
To encode non-persistence in the color of each Gaussian while preserving the appearance of persistent ones, we extend Equation~\ref{equ:multiTimeColor} by introducing the persistence filter $S$, yielding the highlight color~$H$:
\begin{gather}\label{equ:highlightedColor}
    H(t) = \sum_{i \in N} B_i(\hat{t}) \, S_i(t)
        \prod^{i-1}_{j=1}(1-\alpha_j  o_j \cdot \hat{t}) 
        \; \text{.}
\end{gather}
With the persistence filter returning a three-component vector for an individual Gaussian as
\begin{gather}\label{equ:specFilter}
    S_i(t) = \sum_{t_p \in t} \Big(
        t_p (o^p_i + 1 - \max(o_i \odot t)) \int_{T_{p}}^{T_{p+1}} \lambda(x) \; dx
    \Big) 
    \; \text{,}
\end{gather}
with $T_p = \sum_{k=0}^{p-1} \hat{t}_k$. 
$\lambda$ is a vector-valued function mapping the opacity manipulation to a three-component vector ($\lambda: \mathbb{R} \to \mathbb{R}^3$), consisting of a shifted Gaussian function, defined as 
\begin{align}\label{equ:gaussFilterFunctions}
    \lambda(x) =
    \begin{bmatrix}
    \frac{2}{\sigma\sqrt{2\pi}} \exp\left( -\frac{x^2}{2\sigma^2} \right) \\
    \frac{1}{\sigma\sqrt{2\pi}} \exp\left( -\frac{(x-0.5)^2}{2\sigma^2} \right) \\
    \frac{2}{\sigma\sqrt{2\pi}} \exp\left( -\frac{(x-1)^2}{2\sigma^2} \right)
    \end{bmatrix} \text{.}
\end{align}
When integrated, this results in $\int_0^1 \lambda(x) \, dx \approx  [1,1,1]^T$. 
Therefore, in the case of a Gaussian that is persistent over the whole time selection, we have $S \approx [1, 1,1]^T$, whereas a Gaussian that is not visible at the beginning of the selected time period would result in a vector $\lambda(x)$ with the first component close to 0. 
Since the persistence of a Gaussian is encoded by the relative differences between 
the components of $o_i$ rather than their absolute values, 
we shift $o_i$ by $1 - \max(o_i \odot t)$. This ensures that a Gaussian with $\max(o_i \odot t) < 1$ but with $o_i^p \approx o_i^q$ for all $p$ and $q$ still results in $S \approx [1,1,1]^T$, indicating a persistent Gaussian.
Figures~\ref{fig:total_combined_figure} and \ref{fig:vases_all_times} show results rendered with our change-aware approach as well as the blended true-color version. 
Figure~\ref{fig:result_synthetic} shows the synthetic coral dataset where whole objects are added and removed across timesteps. 
Figure~\ref{fig:time_selection_highlight_little} shows a time selection with little change between the first two timesteps, resulting in a nearly unaltered representation, while Figure~\ref{fig:time_selection_highlight_all} extends the selection to a third timestep that differs significantly, producing more pronounced color highlights.
}


\subsection{Increasing Visual Distance}
{
To make the change visualization stand out more clearly and to separate the change highlight visualization and the true colors of the Gaussians, the color $B$ can be altered to grayscale $B_{\textsc{gray}}$, or the contrast can be decreased. Depending on the use case, either a clear highlight of the changes, or the preservation of the overall visual representation is desired.
Since $o_i^p$ is continuous, there is no discrete threshold separating persistent 
from non-persistent Gaussians. To control the visual contrast of the change 
highlight, we optionally apply the parametric sigmoid
\begin{gather}
\phi(x) = \frac{x^k}{x^k + \left(\frac{1-b}{b} (1-x)\right)^k}
\end{gather}
to $o_i^p$ in Equation~\ref{equ:specFilter}, where $k$ controls the contrast 
between persistent and non-persistent Gaussians and $b$ controls the midpoint 
of the transition.
}

\begin{figure}[t]
    \centering
    \begin{subfigure}[b]{0.49\columnwidth}
        \centering
        \includegraphics[width=\textwidth]{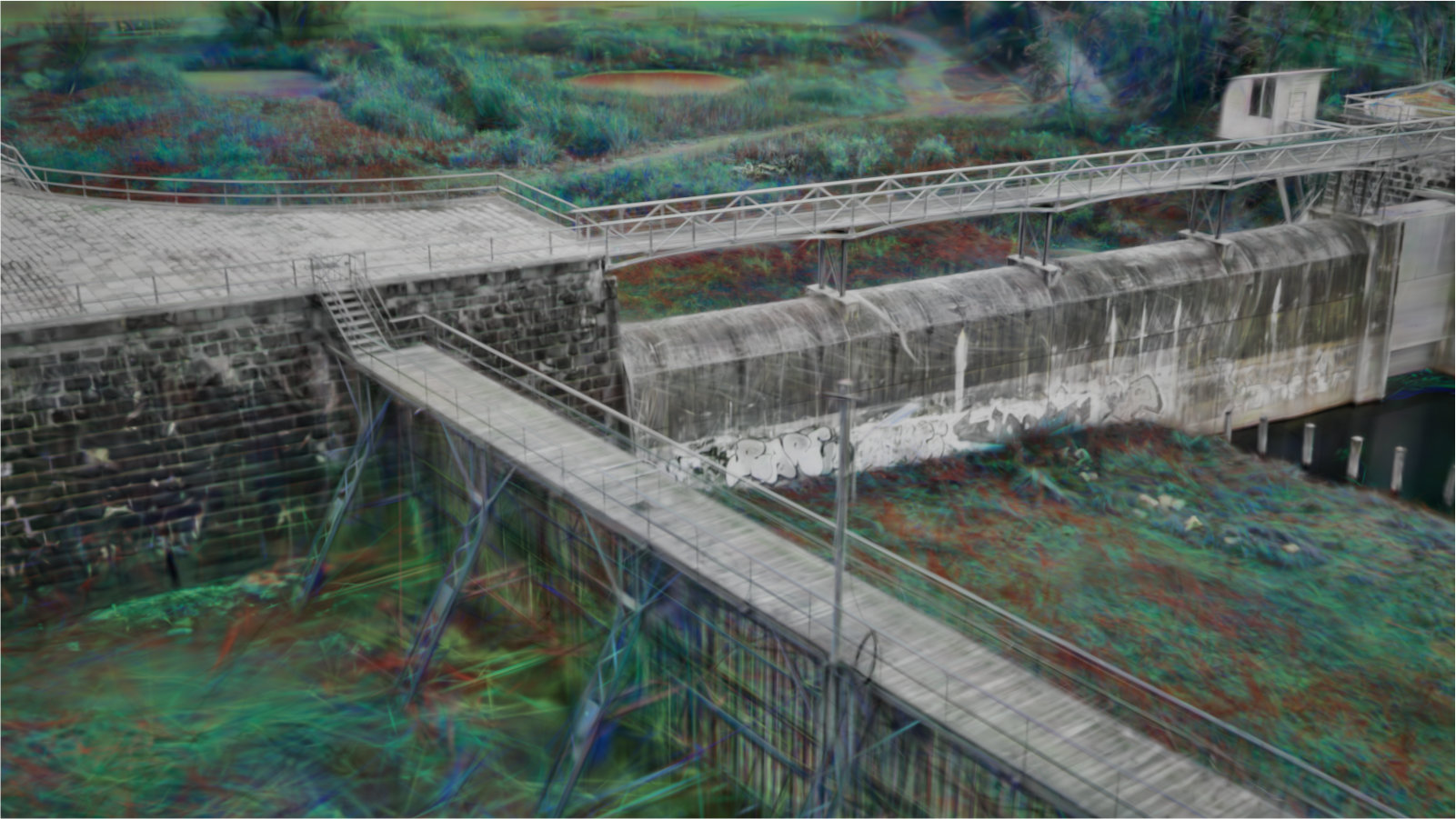}
        \caption{Full RGB color space}
        \label{fig:grid_a}
    \end{subfigure}
    \begin{subfigure}[b]{0.49\columnwidth}
        \centering
        \includegraphics[width=\textwidth]{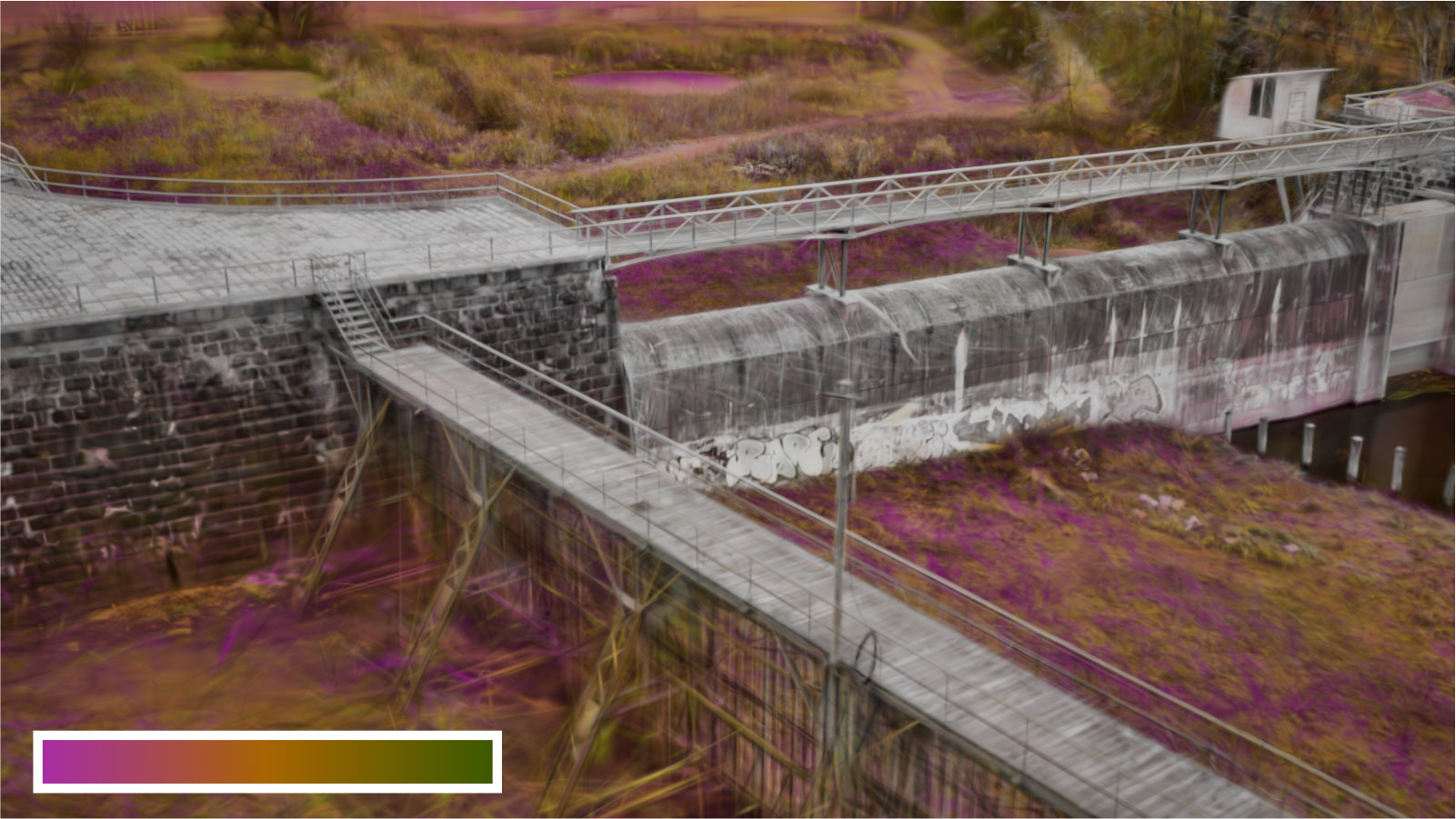}
        \caption{Three color gradient}
        \label{fig:grid_b}
    \end{subfigure}
    \begin{subfigure}[b]{0.49\columnwidth}
        \centering
        \includegraphics[width=\textwidth]{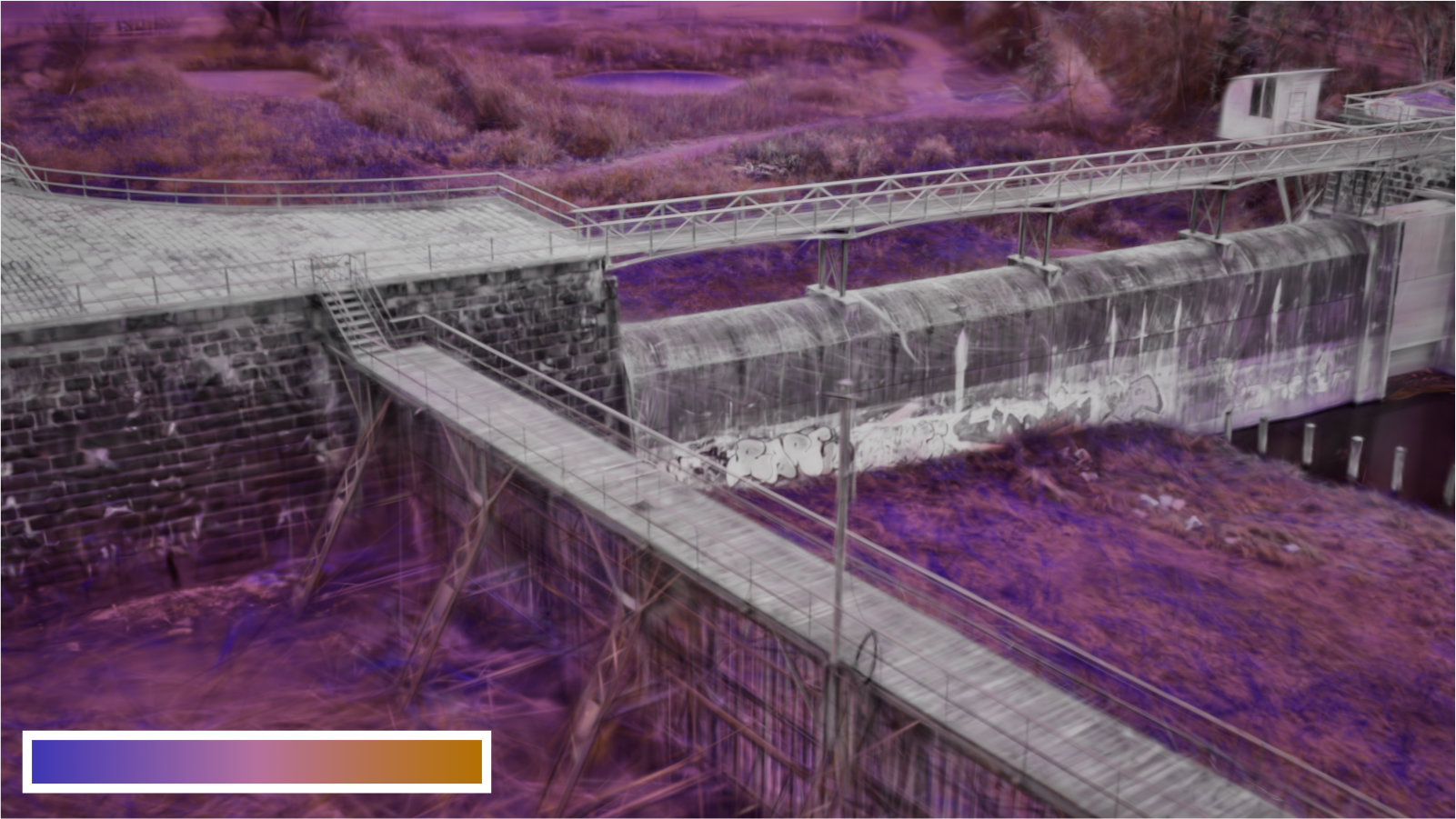}
        \caption{Three color gradient}
        \label{fig:grid_c}
    \end{subfigure} 
    \begin{subfigure}[b]{0.49\columnwidth}
        \centering
        \includegraphics[width=\textwidth]{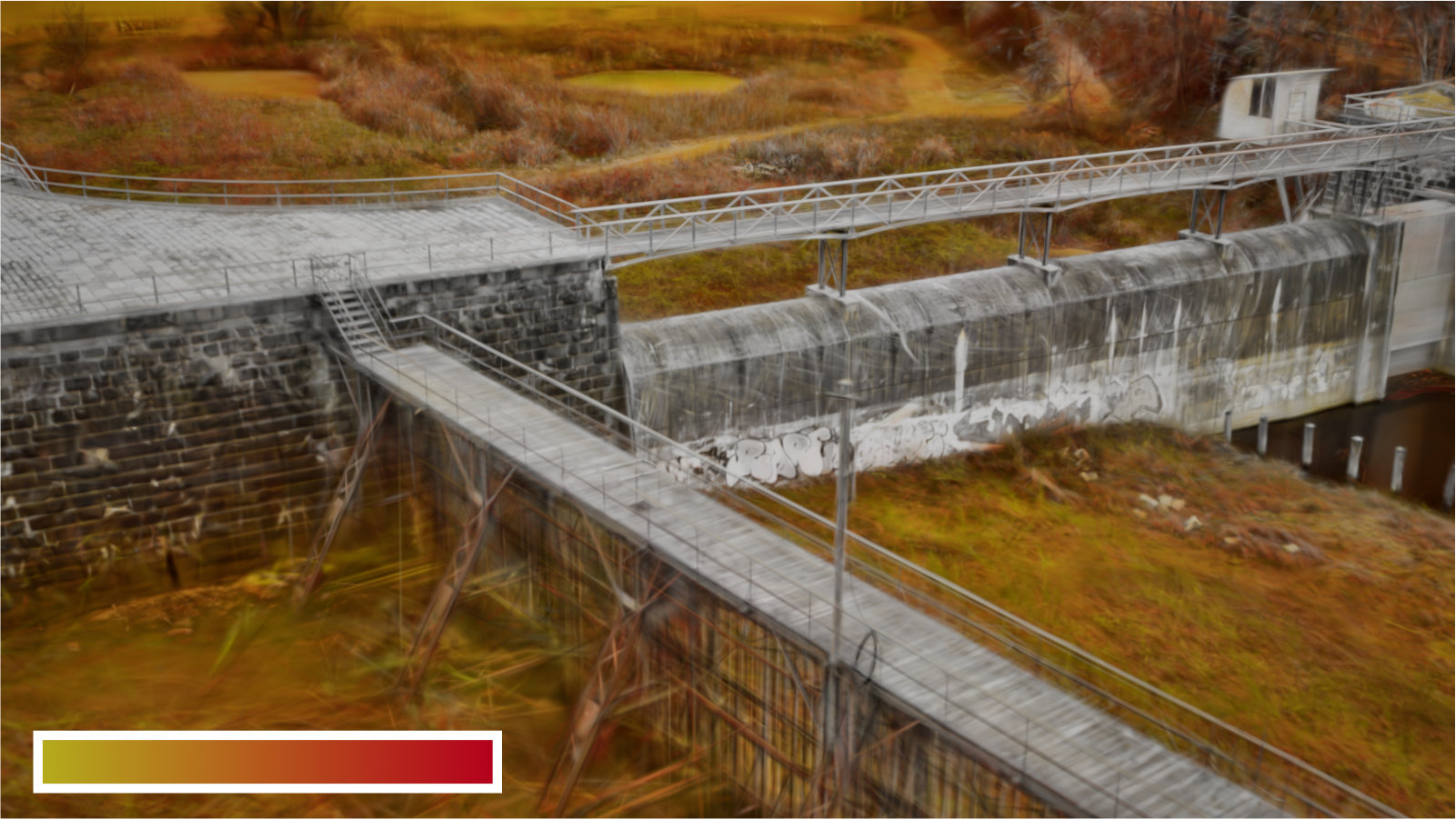}
        \caption{Two color gradient}
        \label{fig:grid_d}
    \end{subfigure}   
    \caption{Different gradients for change highlighting across four selected timesteps.}
    \label{fig:different_gradient}
\end{figure}

\subsection{Custom Highlight Colors}
The persistence filter $S$, given in Equation~\ref{equ:specFilter}, maps the time of changes to the full RGB space, where the direction of $S_i$ encodes when a change occurred, and its magnitude encodes the overall contribution of the filter. For visualization purposes, we replace the RGB hue space with a custom gradient $\gamma$, defined by colors $v = [v_0, v_1, \dots]$ evenly spaced along the gradient with linear interpolation between adjacent pairs, with $\gamma(\rho)$ denoting the color at position $\rho$. This remapping replaces the directional component of $S_i$ with the corresponding gradient color while preserving $\|S_i\|_1$ and the saturation of $S_i$.

{
To remap $S_i$ to a custom gradient $\gamma$, we re-express its directional component in terms of $\gamma$ while retaining its magnitude and saturation.
First we compute the scalar position $\rho \in [0, N{-}1]$ into the $N$-color
gradient by normalizing $S_i$ and projecting onto the temporal axis $\rho = (\hat{S}_i^2 / 2 + \hat{S}_i^3)(N-1)$,
where $\hat{S}_i = S_i / \|S_i\|_1$, and $\hat{S}_i^2$ and $\hat{S}_i^3$ are the second and third components of $\hat{S}_i$ respectively.
The gradient color $\tilde{\gamma}$ is then obtained by linearly interpolating between the two adjacent gradient samples $v_{\lfloor\rho\rfloor}$ and $v_{\lfloor\rho\rfloor+1}$.
We rescale the sampled gradient color by the ratio of input to target magnitude
\begin{align}
    \tilde{S}_i = \tilde{\gamma}(\rho) 
            \frac{\lVert S_i \rVert_1}{
                \lVert  \tilde{\gamma}(\rho) \rVert_1
            }
\end{align}
and scale the saturation of $\tilde{S}_i$ by the saturation of $\hat{S}_i$ in HSV color space.
Together, the magnitude rescaling and saturation scaling ensure that a fully persistent Gaussian ($S_i \approx [1,1,1]^T$) produces $S'_i \approx [1,1,1]^T$
regardless of the chosen gradient, leaving the Gaussian's color unaffected. An example of different gradient colors can be seen in Figure~\ref{fig:different_gradient}.
}

\section{Implementation Details}\label{sec:imp_details}

Our approach is independent of the training method of the initial Gaussian Splatting sets that represent a single timestep. For large scenes, where the number of Gaussians is critical for real-time rendering, we train our initial Gaussian Splatting sets using PUP 3D-GS~\cite{HansonTuPUP3DGS}. For small-scale scenes, where we aimed to preserve finer details, we used the original Gaussian Splatting implementation~\cite{kerbl2023gaussiansplatting}. For both cases, we limited the training to pre-computed colors and optimized the model with random background colors to prevent background leakage and semi-transparent parts. For large outdoor scenes, we used a position learning rate of $1.6 \times 10^{-5}$, and $1.6 \times 10^{-4}$ for the other test scenes.

For our training procedure, we extended the differentiable renderer by Kerbl et al.~\cite{kerbl2023gaussiansplatting} to include light compensation and opacity manipulation. 
For the cross-timestep initialization, we run each Gaussian set against each other ground-truth image set for $3000$ iterations. We only optimize $o$ and $l$ in this run, using a learning rate of $0.0075$ for $o$ and $0.00375$ for $l$. We start with randomly colored backgrounds and gradually limit the background color to black. 
To restrict components of $o$ and $\alpha$ to the interval $(0, 1)$, we activate the values with a sigmoid function. For components of $l$, we aim for an interval of $(0, 2)$, whereas $1$ results in no color shift. Hence, we activate the components of $l$ with $2\sigma(l)$.

For our combined optimization run, we optimize all parameters of the Gaussians except $l$. In our experiments, there was little improvement in optimizing $l$ further, since the cross-timestep initialization should already result in a good light compensation. We use an experimentally determined learning rate of $2.5 \times 10^{-6}$ for $o$, $0.005$ for $\alpha$, $2 \times 10^{-5}$ for $c$, the same learning rate for the position parameter as when training the input Gaussian sets, and $5 \times 10^{-5}$ for the remaining parameters. We use a smaller learning rate for $o$ than for $\alpha$ to preserve a strong encoding of the persistence of the Gaussians.
In every training iteration, we select a training image from across all timesteps.
 At regular intervals during refinement, we check, for each Gaussian, whether it contributes to any training image of $I^p$ when rendered from the corresponding training cameras. If a Gaussian is never rendered for any image of $I^p$, no gradient flows to its time-dependent parameters for that timestep, leaving $o_i^p$ unconstrained. We therefore explicitly set $o_i^p = 0$ in this case. After refinement, we prune all Gaussians where $\alpha_i o_i^p \leq 1/255$ for all $p$, removing Gaussians that are never visible across any timestep.

We implemented our Interactive Renderer for the multi-temporal visualization and change highlighting in Unreal 5.6, using a GPU Niagara Particle System, where each Gaussian splat is represented by a particle sprite facing the camera. \added{All renderings and FPS measurements are performed at the highest material settings.} We allow user-defined time selections $t$, where each component of $t$ can be 0 or 1. This enables arbitrary time selections, for example, to render the scene and highlight the changes for $t_1$ and $t_4$.
In our implementation, we transform the Gaussian color $c$ to the linear color space and perform all computations there.  Our code and dataset \replaced{are publicly available at \href{https://tobiasbat.github.io/ChronoFuseGS/}{https://tobiasbat.github.io/ChronoFuseGS/}}{will be made public upon acceptance}.

\begin{table}[t]
    \centering
    \footnotesize
    \begin{tabularx}{\columnwidth}{Xccc}
        Name & T & Days & Images \\
        \midrule
        No Wolf in the Meadow (NWM)  & & & \\
        \hspace{5mm} \textit{Complete} & 9 & 8 & 2221 \\
        \hspace{5mm} \textit{Autumn} & 4 & 4 & 975 \\
        \hspace{5mm} \textit{Flooding} & 3 & 3 & 629 \\
        \hspace{5mm} \textit{Snow}  & 3 & 1 & 608 \\
        \hspace{5mm} \textit{Car} & 3 & 1 & 705 \\
        Baluster Vase & 4 & - & 200 \\
        Corals & 4 & - & 300 \\
        \bottomrule
    \end{tabularx}
    \caption{Overview of datasets, including the number of timesteps T, recording days, and total input images. The snow subset includes two timesteps that are not present in the complete dataset. Some timesteps are used for multiple subsets. Synthetic datasets (Baluster Vase, Corals) have no associated recording days.}
    \label{tab:datasets}
\end{table}

\begin{figure*}[t]
    \centering
    \includegraphics[width=\textwidth]{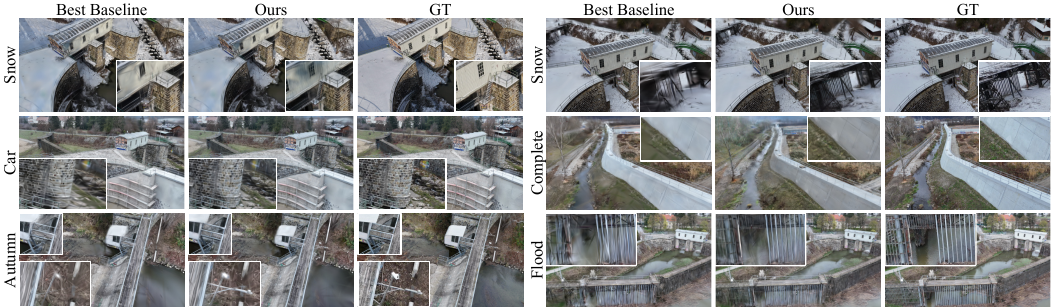}
    \caption{Reconstruction results on five subsets of the No Wolf in the Meadow dataset. Rows correspond to dataset subsets; columns show the best-performing individually trained baseline, \mbox{ChronoFuseGS} at 30k refinement iterations, and the ground-truth test image.}
    \label{fig:eval_details}
\end{figure*}

\begin{figure}[t]
    \centering
    \begin{subfigure}[b]{0.495\columnwidth}
        \centering
        \includegraphics[width=\textwidth]{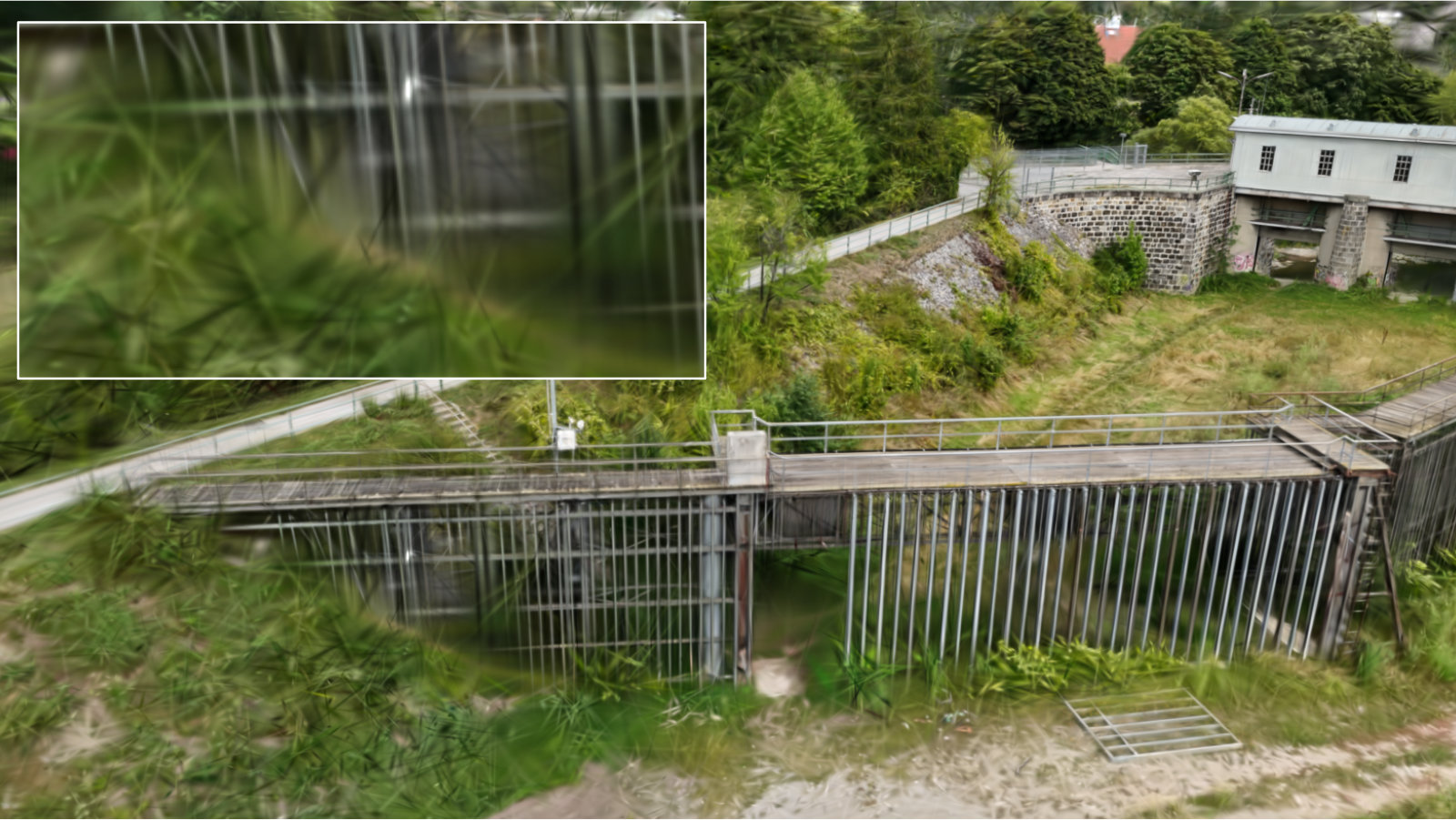}
        \caption{Autumn}
    \end{subfigure}
    \hfill
    \begin{subfigure}[b]{0.495\columnwidth}
        \centering
        \includegraphics[width=\textwidth]{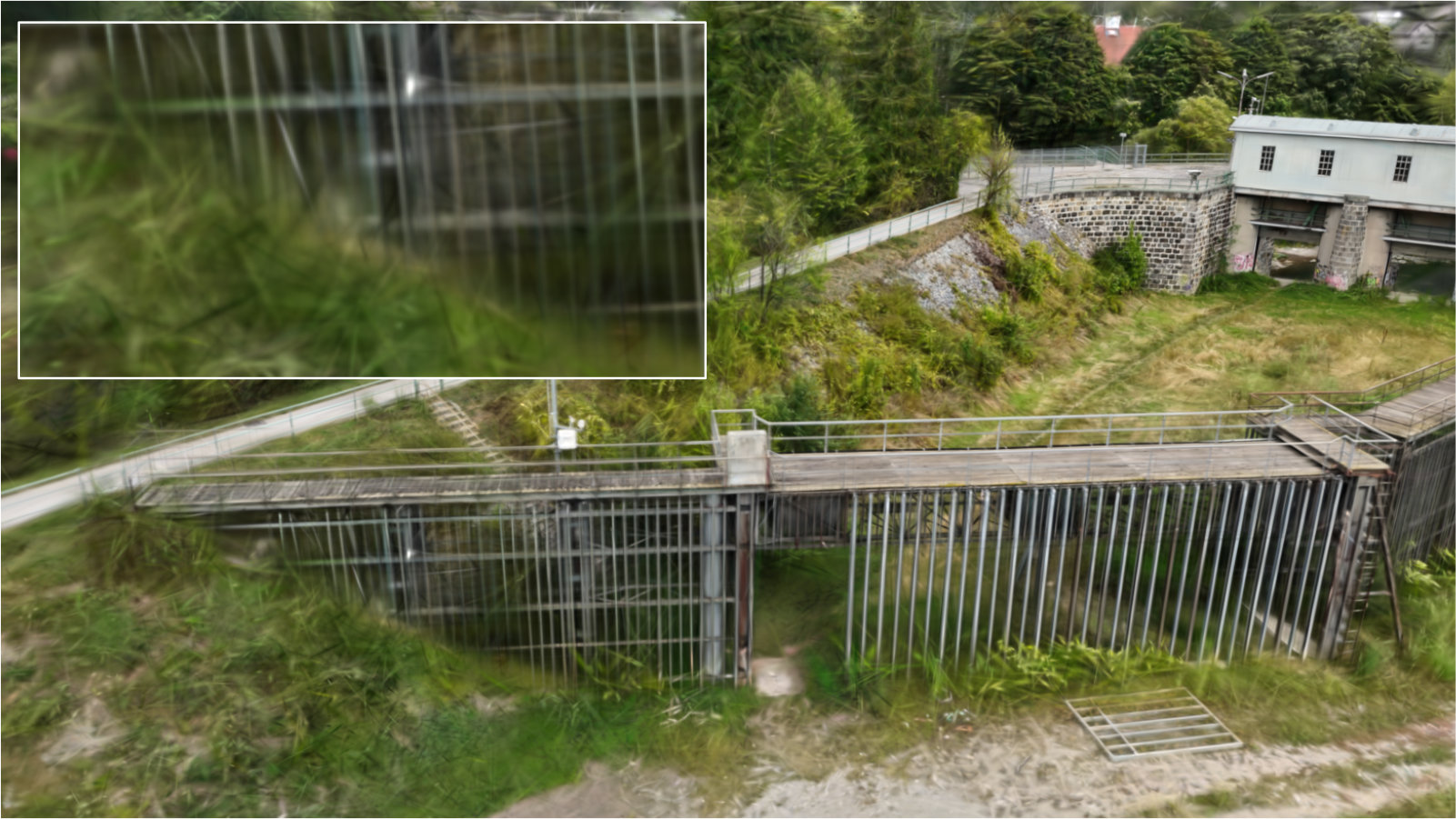}
        \caption{Complete}
    \end{subfigure}
    \caption{Results of our approach with ${30,000}$ refinement iterations for different datasets, where (a) and (b) share the shown timestep. The Autumn subset includes 4 timesteps, and the Complete NWM dataset includes 9 timesteps.}
    \label{fig:result_complete_vs_autom}
\end{figure}

\section{Dataset}
To evaluate our approach, we captured a real-world outdoor dataset spanning multiple months, named \textit{No Wolf in the Meadow (NWM)}. 
The dataset captures a flood control system in \replaced{Vienna, Austria}{ANONYMIZED CITY}, covering an area of approximately $100 \times 200$ meters with a mix of natural and artificial elements, including weir and flood control structures constructed from stone, wood, steel, and concrete, multiple river branches, and vegetation. We captured the area using a consumer drone with pre-recorded flight plans, recording 4K video and exporting individual frames at regular intervals, downscaled to $1600 \times 900$ pixels. 
The dataset additionally includes detailed close-up recordings of specific parts of the scene. 
In total, we captured the area on eight days, with the first recording on 25th August 2025 and the last on 28th March 2026. Beyond seasonal vegetation changes and varying lighting conditions, the dataset includes days when the area was covered by snow or flooded, as well as periods during which reinforcement and renovation work were carried out.
Besides the full dataset covering all eight days, we created subsets focusing on specific types of change -- flooding, snow, and vegetation change -- to simulate a more realistic data collection process, for example, during a natural disaster.
A list of all datasets and their input image counts is provided in Table~\ref{tab:datasets}.

We additionally created two synthetic datasets, each consisting of 3D-scanned and textured objects~\cite{smithsonian3d} rendered in Blender at $1600 \times 900$ pixels. The Baluster Vase scene features three vases across four simulated timesteps, with new cracks added to the central vase at each timestep. The Coral scene arranges multiple corals and shells that appear and disappear across four simulated timesteps. In both scenes, lighting remains constant across timesteps. Results are shown in Figures~\ref{fig:vases_all_times} and \ref{fig:result_synthetic}.

\definecolor{best}{rgb}{0.70,0.90,0.70}         
\definecolor{second_best}{rgb}{0.88,0.95,0.88}  
\definecolor{third_best}{rgb}{0.94,0.98,0.94}   

\newcommand{\bc}[1]{\cellcolor{best}#1}
\newcommand{\sc}[1]{\cellcolor{second_best}#1}
\newcommand{\tc}[1]{\cellcolor{third_best}#1}
\begin{table*}[h] 
    \centering
    \setlength{\tabcolsep}{7.3pt}
    \footnotesize
    \begin{tabularx}{\textwidth}{l *{15}{c}}
        & \multicolumn{3}{c}{Complete (4.67)} 
        & \multicolumn{3}{c}{Autumn (2.17)} 
        & \multicolumn{3}{c}{Flooding (1.65)} 
        & \multicolumn{3}{c}{Snow (0.81)} 
        & \multicolumn{3}{c}{Car (1.56)} \\
        \cmidrule(lr){2-4} \cmidrule(lr){5-7} \cmidrule(lr){8-10} 
        \cmidrule(lr){11-13} \cmidrule(lr){14-16}
        Method 
        & PS$\uparrow$ & SS$\uparrow$ & LP$\downarrow$
        & PS$\uparrow$ & SS$\uparrow$ & LP$\downarrow$
        & PS$\uparrow$ & SS$\uparrow$ & LP$\downarrow$
        & PS$\uparrow$ & SS$\uparrow$ & LP$\downarrow$
        & PS$\uparrow$ & SS$\uparrow$ & LP$\downarrow$ \\
        \midrule
        Individual Init      & 21.6 & 0.73 & 0.45   & 20.9 & 0.69 & 0.47   & 22.2 & 0.76 & 0.43   & 23.2 & 0.82 & 0.38   & 22.7 & 0.76 & 0.44 \\
        Ours Merged     & 21.6 & 0.73 & 0.45   & 20.9 & 0.69 & 0.47   & 22.2 & 0.76 & 0.43   & 23.2 & 0.82 & 0.38   & 22.7 & 0.76 & 0.44 \vspace{2mm}\\
        Ours 6k         & 21.5 & 0.71 & \tc0.44   & \sc21.2 & \sc0.70 & \sc0.46   & \sc22.8 & \tc0.78 & \tc0.41   & \tc23.3 & \tc0.82 & \tc0.37   & \tc23.1 & \tc0.78 & \tc0.42 \\
        Ours 15k        & \sc21.8 & \tc0.73 & \sc0.43   & \tc21.1 & 0.69 & \tc0.47   & \tc22.8 & \sc0.78 & \sc0.41   & \sc23.4 & 0.82 & \sc0.36   & \sc23.2 & \sc0.78 & \sc0.42 \\
        Ours 30k        & \bc21.9 & \bc0.74 & \bc0.43   & \bc21.3 & \bc0.70 & \bc0.46   & \bc22.8 & \bc0.78 & \bc0.40   & \bc23.4 & \bc0.83 & \bc0.36   & \bc23.3 & \bc0.79 & \bc0.41 \vspace{2mm}\\
        Individual +6k  & 21.5 & 0.73 & 0.45   & 20.7 & 0.67 & 0.48   & 21.7 & 0.73 & 0.44   & 23.2 & 0.82 & 0.38   & 22.3 & 0.73 & 0.45 \\
        Individual +15k & \tc21.6 & \sc0.73 & 0.45   & 21.0 & \tc0.69 & 0.47   & 22.1 & 0.76 & 0.43   & 23.2 & \sc0.82 & 0.37   & 22.7 & 0.76 & 0.44 \\
        Individual +30k & 21.4 & 0.73 & 0.45   & 20.6 & 0.68 & 0.47   & 21.8 & 0.73 & 0.46   & 23.1 & 0.82 & 0.37   & 22.5 & 0.76 & 0.44 \\
        \bottomrule
    \end{tabularx}
    \caption{Quantitative comparison across the five subsets of the No Wolf in the Meadow dataset. Numbers in parentheses indicate the combined number of Gaussians of the input Gaussian sets in millions. PS: PSNR, SS: SSIM, LP: LPIPS. $\uparrow$ indicates higher is better, $\downarrow$ lower is better. \colorbox{best}{Best}, \colorbox{second_best}{second best}, and \colorbox{third_best}{third best} results are highlighted per column. Individual Init represents the individual trained Gaussian sets used as input for our approach, and Ours Merged represents the multi-temporal model after merging the individual Gaussian sets before additional training. }
    \label{tab:simularity_scores}
\end{table*}

\section{Evaluation}

\subsection{Reconstruction Quality}
\label{sec:results}
{
We evaluated whether our multi-temporal Gaussian fusion approach achieves better generalization compared to individually trained Gaussian sets. In particular, we are interested in whether Gaussians detected as persistent by our system and originating from one timestep also contribute to the reconstruction at other timesteps. Additionally, we evaluated whether our merging process~-- the computation and transformation into a shared coordinate system~-- is sufficiently precise and robust for real-world data.
}

{
For our experiments, we train each timestep individually using PUP 3D-GS~\cite{HansonTuPUP3DGS}. We merge the individual Gaussian sets and refine the resulting multi-temporal model for $6{,}000$, $15{,}000$, and $30{,}000$ iterations, in addition to the initialization step, using the same parameters and learning rates as outlined in Section~\ref{sec:imp_details}. As a baseline, we compare against individually trained Gaussian sets with additional iterations ($6{,}000$, $15{,}000$, and $30{,}000$) for the initial training, using the same parameters as the individual sets that serve as input to our approach. We split all datasets into training and test sets, where every 8th image from the ordered list of images for each individual timestep serves as a test image. We evaluate on the five subsets of the No Wolf in the Meadow dataset, which cover a range of scene-change types, including seasonal vegetation, snow, and flooding.
}

{
We evaluated each method at each saved iteration checkpoint using the average PSNR, SSIM~\cite{wang2004image}, and LPIPS~\cite{zhang2018perceptual} over the test split. For our method, we additionally report results for the initial merged model prior to any refinement. For the individually trained baselines, metrics are averaged across timesteps weighted by the number of images per timestep. Additionally, we manually compare and analyze the representation of fine details in the individual novel view synthesis. 
}

{
The results of the quantitative analysis are listed in Table~\ref{tab:simularity_scores}. As expected, the individually trained models serving as input to our approach, and the multi-temporal model after the merging stage, achieved equivalent scores across all datasets. 
Across all five datasets and three metrics, our approach consistently outperformed the individually trained baselines, with Ours 30k achieving the highest scores in all cases. Our approach, with the fewest additional iterations (6k), already outperformed the best individual baseline across all datasets in terms of LPIPS, with improvements or comparable results in PSNR and SSIM. For the individual baselines, differences across iteration counts were minimal, with some +30k scores performing worse than +6k (e.g. Flooding LPIPS: +6k = 0.44, +30k = 0.46). Within our approach, scores generally improved with more iterations, though the gains were marginal. Overall, while our approach outperformed the individually trained models, the differences in PSNR, SSIM, and LPIPS are small, as these metrics average over the full image and do not capture fine-detail improvements well.
For fine-grained details, our approach yields greater improvements than additional iterations of individually trained models, recovering whole structures absent in single-timestep reconstructions and noticeably increasing reconstruction detail (Figure~\ref{fig:eval_details}).
}

\added{
To validate the joint optimization of $\alpha_i$ and $o_i$ during the combined refinement stage, we ran two ablations on the Autumn dataset: one freezing $\alpha_i$ 
and one freezing $o_i$ during refinement (6{,}000 iterations). Both performed worse than our combined optimization (PSNR/SSIM/LPIPS: 21.2/0.70/0.46): freezing $o_i$ yields 21.0/0.68/0.47 and freezing $\alpha_i$ yields 20.2/0.66/0.49.
}

We observed that adding more timesteps to the fused reconstruction increases the level of detail across all timesteps. As shown in Figure~\ref{fig:result_complete_vs_autom}, the complete reconstruction contains more detail than the subset, even though both models were refined for the same number of iterations -- meaning the multi-temporal model received fewer iterations per timestep. Together with the recorded similarity scores, this suggests that Gaussians are reused across timesteps rather than each timestep being refined independently. 
The number of Gaussians remains stable across methods, with a variation below 
3.2\% across all datasets and iteration counts (exception: Flooding Individual 
+30k, $-$13\%), indicating that the additional detail is not simply a result of a higher total Gaussian count.
Our experiments also highlighted a limitation of our approach: it relies on precise camera parameter estimation through SfM when merging individual models. In the complete dataset results, the camera parameters for the snow timestep were insufficiently precise, resulting in blurry details and edges in the reconstruction.

\begin{table}[h]
    \centering
    \footnotesize
    \setlength{\tabcolsep}{4pt}
    \begin{tabularx}{\columnwidth}{X @{\hspace{2mm}} S[table-format=2.1] S[table-format=2.1] S[table-format=2.1] S[table-format=3.1] @{\hspace{4mm}} S[table-format=3.1] S[table-format=3.1] S[table-format=3.1]}
        & \multicolumn{4}{c}{Training Time (min)} & \multicolumn{3}{c}{Rendering FPS} \\
        \cmidrule(lr){2-5} \cmidrule(lr){6-8}
        Dataset & {Merge} & {Cross} & {Ref} & {Sum} & {$C$} & {$H$} & {$H_{0,1}$} \\
        \midrule
        NWM & & & & & & & \\
        \hspace{3mm} \textit{Complete}  & 75.4 & 48.3 & 12.0 & 135.7 & {48} & {49} & {49} \\
        \hspace{3mm} \textit{Autumn}    & 19.3 & 14.4 &  4.5 &  38.2 & {120} & {120} & {121} \\
        \hspace{3mm} \textit{Flooding}  & 11.3 &  7.5 &  4.0 &  22.7 & {149} & {151} & {151} \\
        \hspace{3mm} \textit{Snow}      & 15.2 &  6.0 &  3.0 &  24.2 & {231} & {233} & {232} \\
        \hspace{3mm} \textit{Car}       & 13.8 &  7.3 &  3.8 &  24.9 & {163} & {163} & {164} \\
        Baluster Vase                   &  4.9 & 14.5 &  4.5 &  23.9 & {184} & {184} & {184} \\
        Corals                          &  6.0 &  8.2 &  1.7 &  15.8 & {264} & {264} & {264} \\
        \bottomrule
    \end{tabularx}
\caption{\added{Training times in minutes for 6000 refinement iterations and average rendering FPS of our Unreal implementation. Merge, Cross, and Ref denote the pipeline stages: \textbf{Merg}ing to Multi-Temporal Model, \textbf{Cross}-Timestep Initialization, and Combined \textbf{Ref}inement. $C$ and $H$ represent the FPS with all timesteps selected in true \textbf{c}olor and change 
\textbf{h}ighlighting respectively. $H_{0,1}$ represents the FPS for change 
highlighting with only the first two timesteps selected.}}
\label{tab:times_fps}
\end{table}

\begin{figure}[t]
    \centering
    \includegraphics[width=\columnwidth]{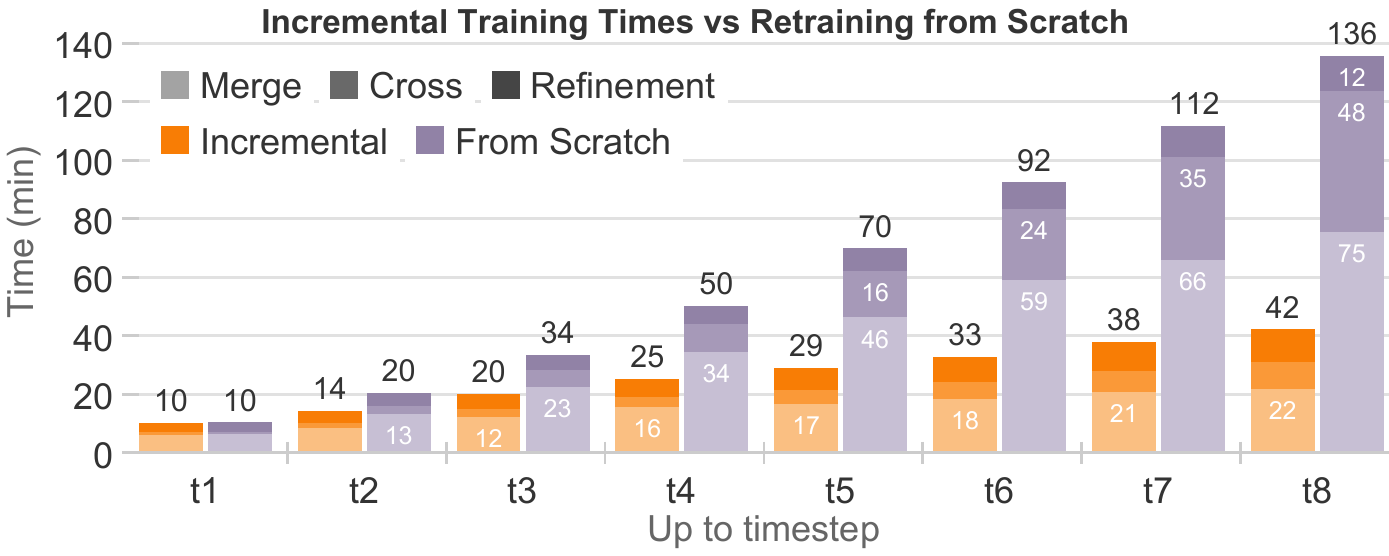}
\caption{\added{Training time up to each timestep of the NWM (complete) dataset, comparing \colorswatch{incrementalcolor} incremental fusion---where the multi-temporal model is extended with a new timestep---against \colorswatch{fromscratchcolor} retraining from scratch, where the model is fully recomputed from all timesteps available so far. Each bar represents the time to incorporate the new timestep and yield a single refined model containing all timesteps from $t_0$ through the new one. Times broken down by pipeline stage: \colorswatch{mergeshade} merging, \colorswatch{crossshade} cross-timestep initialization, and \colorswatch{refinementshade} refinement.}}
\label{fig:time_accumulated}
\end{figure}

\subsection{\added{Training Times \& Rendering Frame Rates}}
\label{sec:training_times}
\added{
Training times and average rendering FPS at a resolution of $1920 \times 1080$ are reported in Table~\ref{tab:times_fps}. Training times and rendering FPS are measured on separate implementations as described in Section~\ref{sec:imp_details}, both on an NVIDIA RTX 4090 with 24GB of VRAM. Figure~\ref{fig:time_accumulated} shows the training times for incrementally adding new timesteps to the NWM (complete) dataset compared to complete recomputation. The incremental approach is significantly faster: adding the last timestep $t_8$ to the model already including $t_0$--$t_7$ required 42 minutes, compared to 136 minutes for a full recomputation of the multi-temporal model, with the merging stage accounting for approximately half of that time in both cases (incremental: 22 min, from scratch: 75 min). This makes the incremental approach particularly well-suited for scenarios where new timesteps become available over time.
}


\subsection{User Study}
{
In addition to the quantitative reconstruction evaluation, we assessed the effectiveness of our visualization method through a within-group user study. We investigated whether users can distinguish between persistent and changing parts of a scene and determine at which timesteps non-persistent parts were present, to what extent interactive time-selection is necessary for this, and for which types of tasks static color highlighting alone is sufficient.
}

We conducted an online survey in which participants were asked to determine the presence of objects at specific timesteps using renderings of a reconstructed scene. We tested three visualization conditions: static color highlighting, interactive color highlighting, and a single-timestep baseline. In the static condition, the scene was rendered using our change highlighting method (Section~\ref{ssec:change_highlight}) with all timesteps selected simultaneously, without the ability to modify the selection. The interactive condition extended the static condition by allowing participants to freely enable and disable individual timesteps, with the highlighting adapting accordingly. All timesteps were selected at the start of each trial. In the baseline condition, participants were presented with renderings of individual timesteps and could navigate between them, with only one timestep visible at a time. An overview of the interface is shown in Figure~\ref{fig:user_interface}. The study comprised 19 tasks, each answered under all three conditions across participants, but each participant completed every task exactly once under a single randomly assigned condition. Tasks were grouped by condition in randomized order, with the condition order also randomized across participants. After each condition block, participants completed a post-exposure questionnaire. There was no time limit, but participants were instructed to be as efficient and accurate as possible.

\begin{figure}[t]
    \centering
    \begin{subfigure}[t]{0.325\columnwidth}
        \centering
        \includegraphics[width=\textwidth]{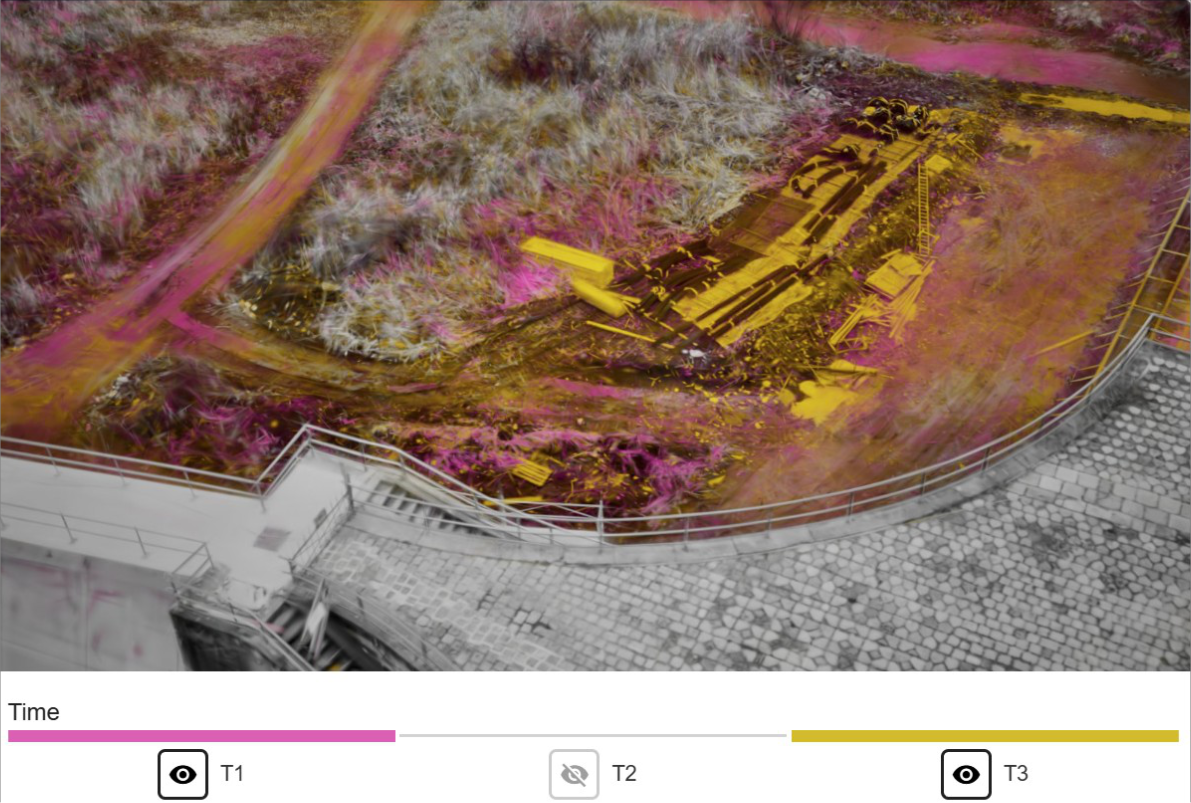}
        \caption{Interactive}
    \end{subfigure}
    \hfill
    \begin{subfigure}[t]{0.325\columnwidth}
        \centering
        \includegraphics[width=\textwidth]{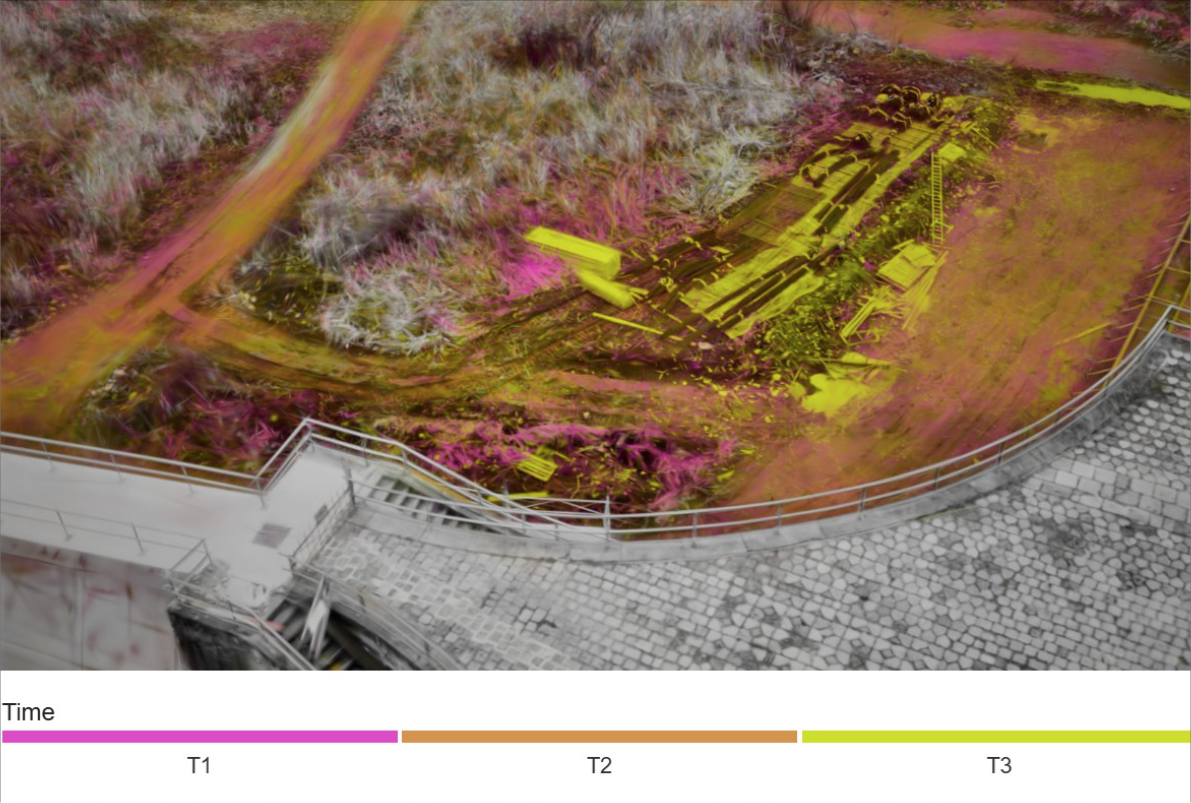}
        \caption{Static}
    \end{subfigure}
    \hfill
    \begin{subfigure}[t]{0.325\columnwidth}
        \centering
        \includegraphics[width=\textwidth]{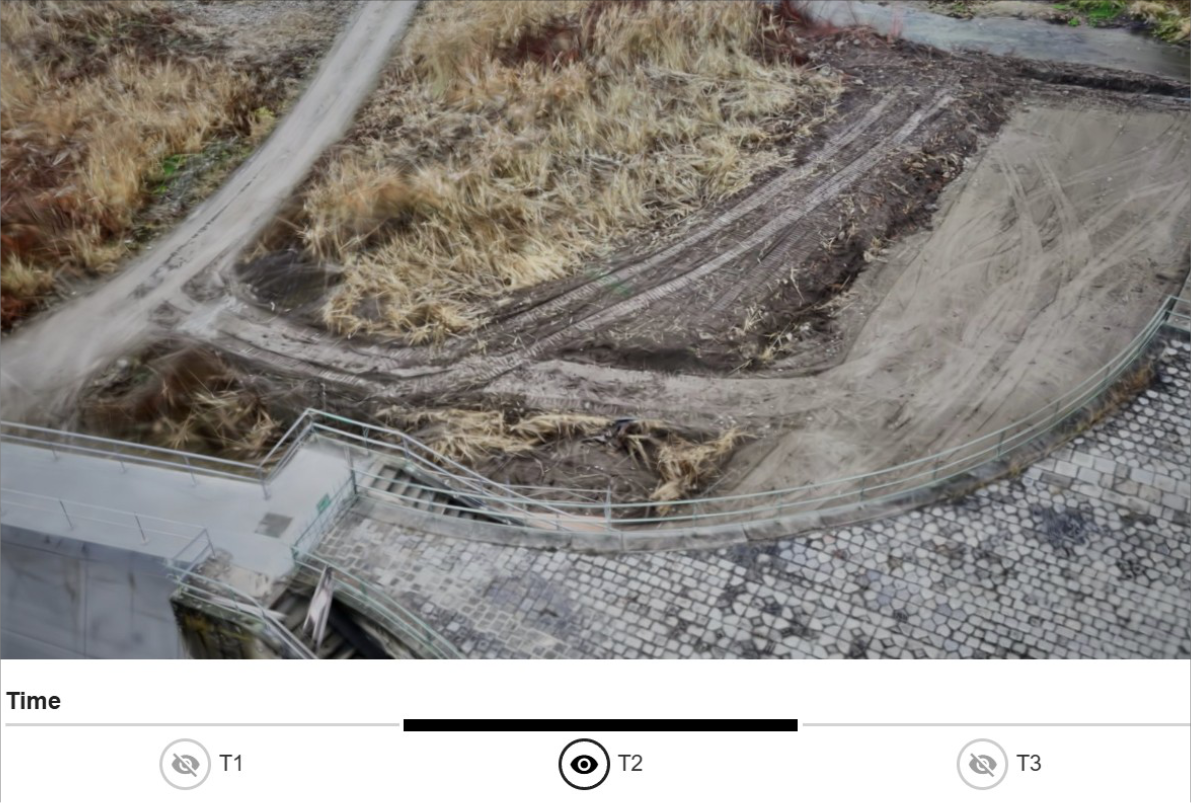}
        \caption{Baseline}
    \end{subfigure}
    \caption{Screenshots of the web survey tool for the three conditions.}
    \label{fig:user_interface}
\end{figure}
\begin{figure}[t]
    \centering
    \begin{subfigure}[t]{0.495\columnwidth}
        \centering
        \includegraphics[width=\textwidth]{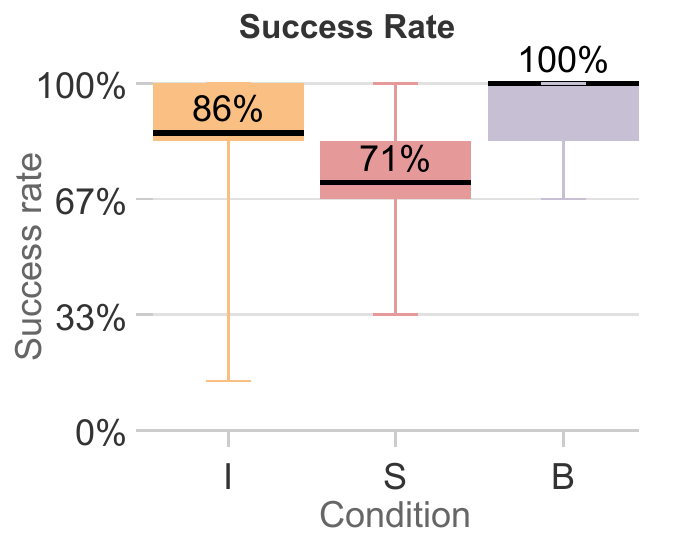}
    \end{subfigure}
    \hfill
    \begin{subfigure}[t]{0.495\columnwidth}
        \centering
        \includegraphics[width=\textwidth]{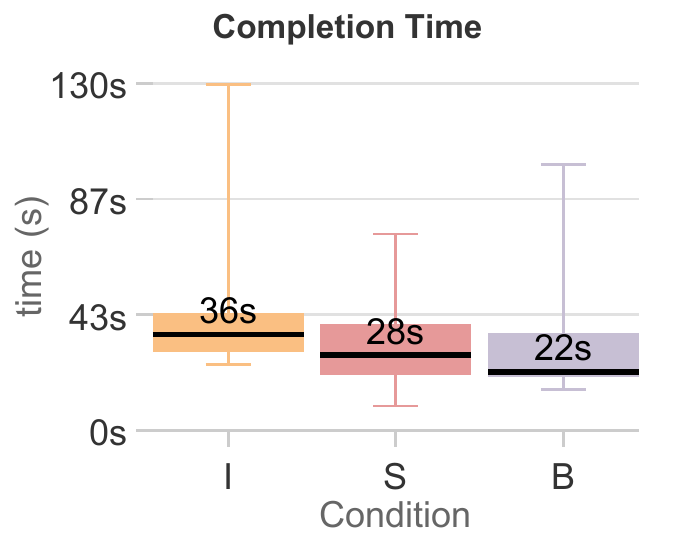}
    \end{subfigure}
    \caption{Quantitative results of the user study across three conditions (I: Interactive, S: Static, B: Baseline). The black line indicates the median.}
    \label{fig:eval_user_results}
\end{figure}

In total, 21 participants completed the study voluntarily, with 11 identifying as male, 8 as female, and 2 not reporting their gender. The average age was 28.7 years (std: 3.81). 14 participants had visualization education, and 10 worked with visualizations professionally.

\begin{figure*}[t]
    \centering
    \begin{subfigure}[b]{0.33\textwidth}
        \centering
        \includegraphics[width=\textwidth]{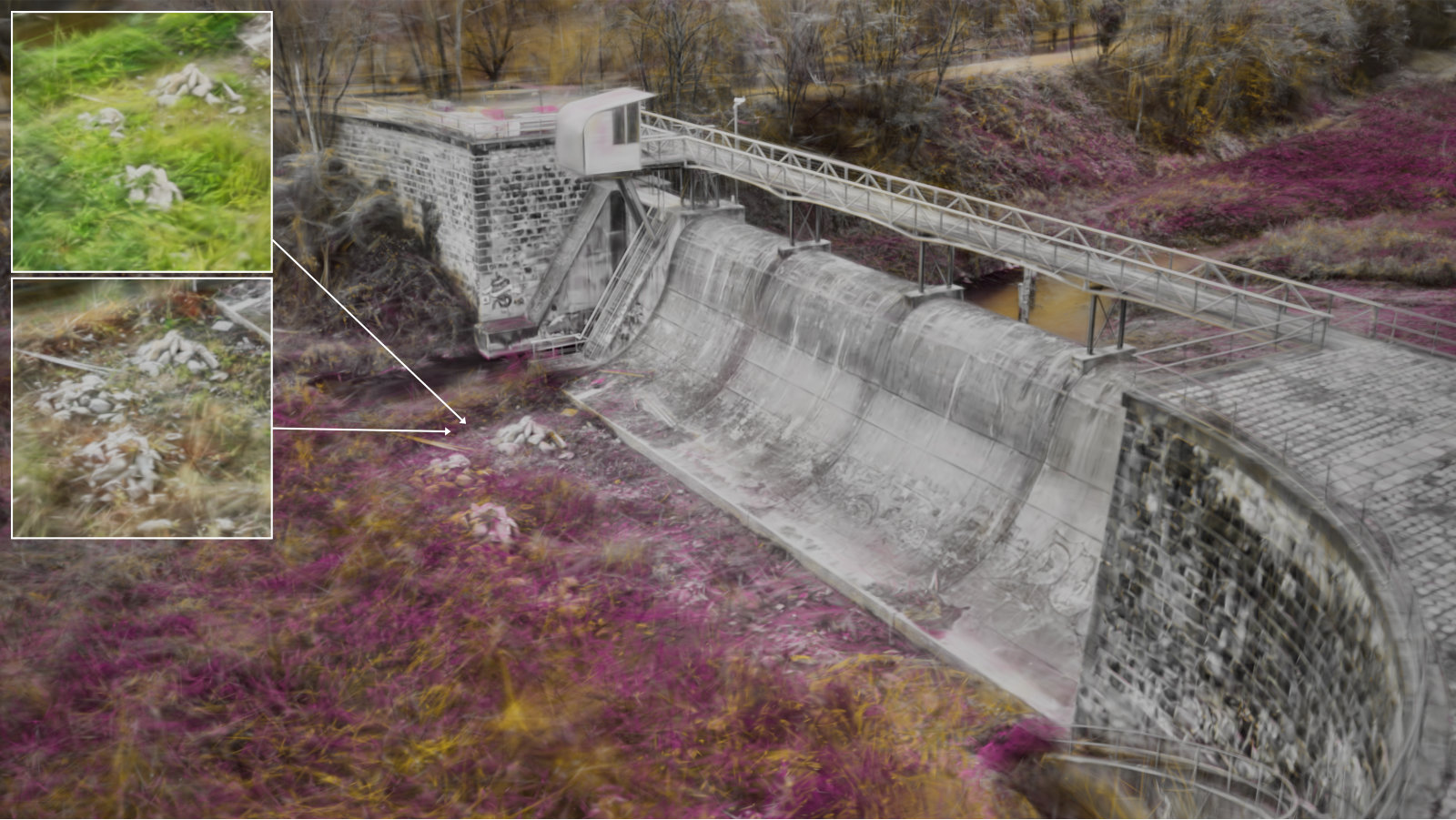}
        \caption{$[$ \tOn[mg_2_1] \tOn[mg_2_2] \tOff \tOff \tOff \tOff \tOff \tOff \tOff$]$}
        \label{fig:highlight_natural}
    \end{subfigure}
    \hfill
    \begin{subfigure}[b]{0.33\textwidth}
        \centering
        \includegraphics[width=\textwidth]{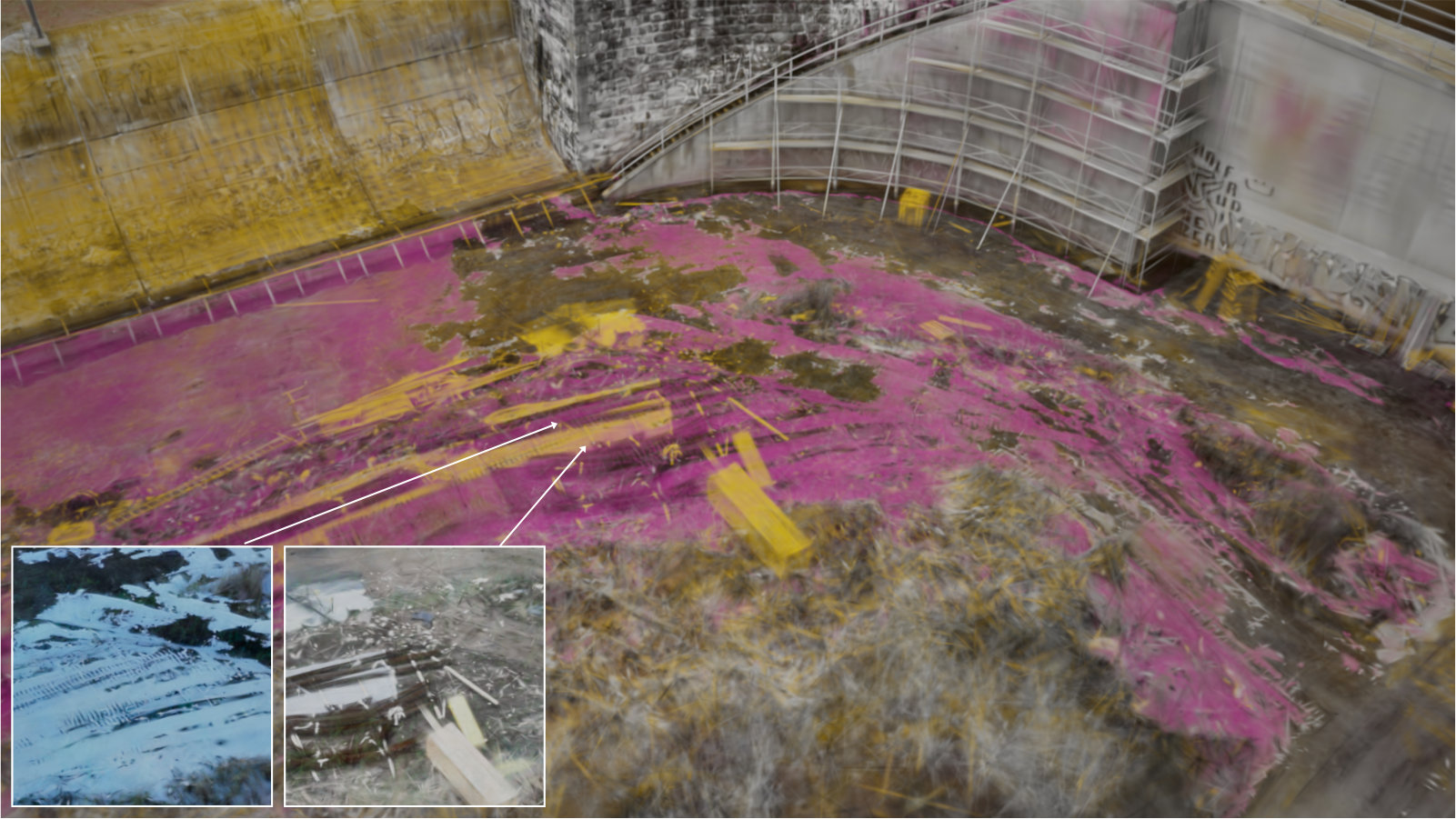}
        \caption{$[$ \tOff \tOff \tOff \tOn[mg_2_1] \tOff \tOff \tOff \tOff \tOn[mg_2_2]$]$}
        \label{fig:highlight_ice}
    \end{subfigure}
        \hfill
    \begin{subfigure}[b]{0.33\textwidth}
        \centering
        \includegraphics[width=\textwidth]{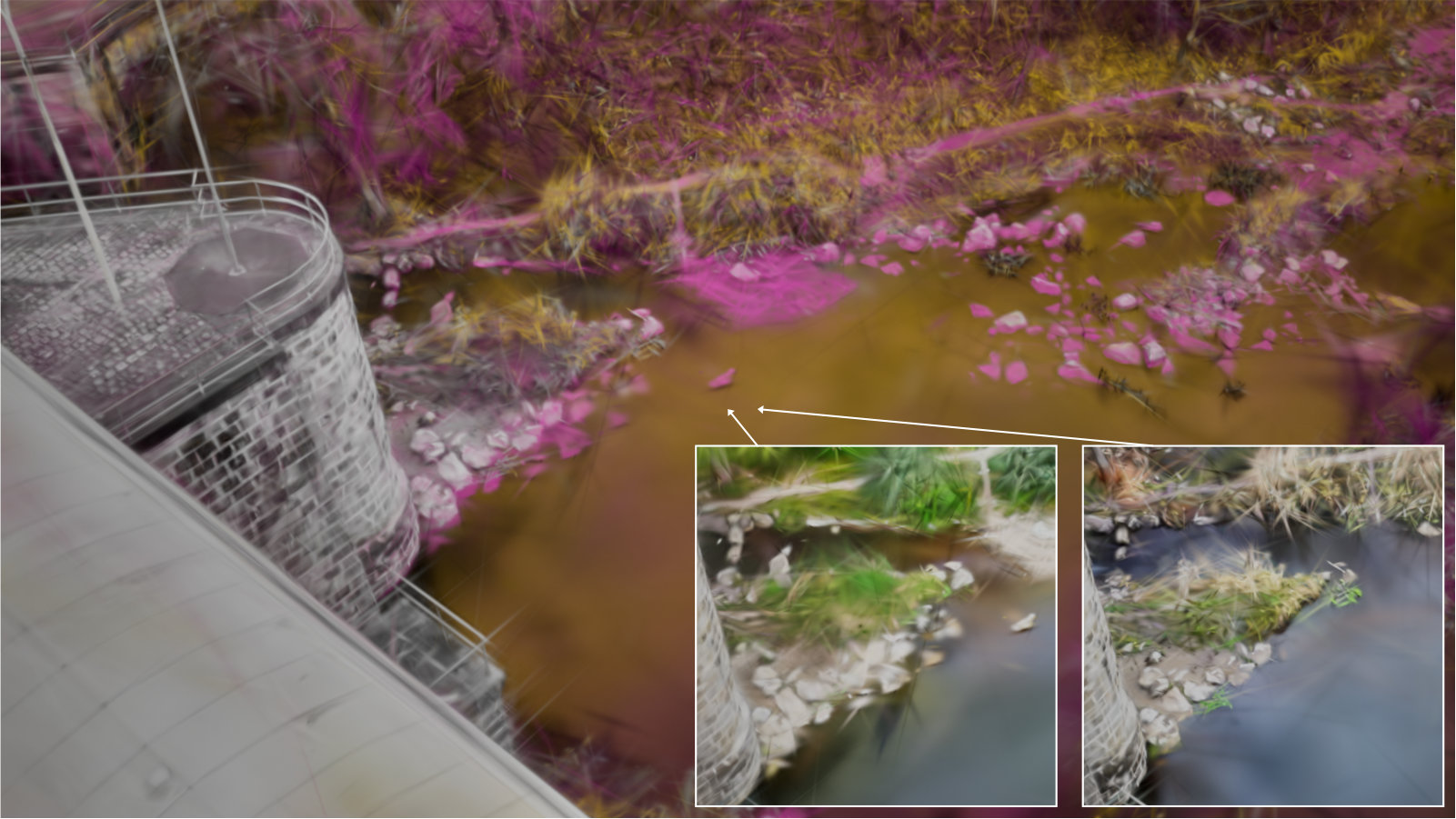}
        \caption{$[$\tOn[mg_2_1] \tOn[mg_2_2] \tOff \tOff$]$}
        \label{fig:highlight_water}
    \end{subfigure}  
  \caption{Change visualization between two timesteps for different scene change types. Insets show the individual timestep renderings. (a) Rocks partially covered by vegetation at different levels. (b) Scene covered with ice at one timestep, with additional objects present during the other timestep. (c) River with differing water levels, highlighting the change in waterline. \tOn indicates an enabled timestep, with color corresponding to timestep presence; \tOff indicates a disabled timestep.}
    \label{fig:highlighting_diff}
\end{figure*}

For each participant and condition, the mean and median success rate and completion time were computed across the completed tasks and then aggregated across all participants. Across all trials, the majority of tasks were answered correctly. The interactive (mean: 83\%, std: 21\%) and baseline conditions (mean: 93\%, std: 10\%) did not differ significantly (Wilcoxon signed-rank, $p=0.12$, $W=32$) with a Bonferroni corrected $\alpha = 0.017$. The static condition performed worst (mean: 72\%, std: 17\%), significantly below both the baseline ($p=0.001$, $W=8$) and interactive condition ($p=0.013$, $W=38$). For completion time, the static (median: 28s, mean: 34s, std: 20s) and baseline conditions (median: 22s, mean: 30s, std: 19s) were not significantly different ($p=0.32$, $W=86$), similarly to the static and interactive conditions ($p=0.035$, $W=55$). The interactive condition (median: 36s, mean: 45s, std: 28s) took significantly longer than the baseline ($p=0.007$, $W=40$). Results are shown in Figure~\ref{fig:eval_user_results}.

When asked how beneficial each method was for quickly spotting changes, participants rated the static method highest (mean 5.6, std 1.2), followed by the interactive method (mean 5.1, std 1.5) and the baseline lowest (mean 5.0, std 1.7), on a seven-point Likert scale from 1 (not at all) to 7 (highly beneficial).
We additionally recorded the time-selection changes performed by participants during each trial of the interactive color highlighting condition. Participants frequently selected three or four timesteps simultaneously, and at the point of answer submission, all timesteps were selected in 60\% of three-timestep tasks and three or four timesteps in 56\% of four-timestep tasks. 
In follow-up interviews, multiple participants noted that both color highlighting conditions required a familiarization period, but that tasks felt easier to solve once they had become accustomed to the approach. Participants also reported that the color highlighting provided a good overview of what had changed and approximately when, but that decomposing the mixed highlight color into individual timesteps was difficult when objects were present across multiple timesteps, requiring additional interaction. 
For example, in one task where only the first timestep was correct, but the majority of participants selected the first two timesteps, indicating that they could approximate when the change occurred, but struggled to decompose the mixed color into individual timesteps. For tasks where users observed a persistent area, the static method performed on par with the other two methods, suggesting that color highlighting alone may be sufficient to identify persistent parts without additional interaction. 
Post-exposure interviews support this observation.

\begin{figure*}[h]
        \begin{subfigure}[b]{0.245\textwidth}
        \centering
        \includegraphics[width=\textwidth]{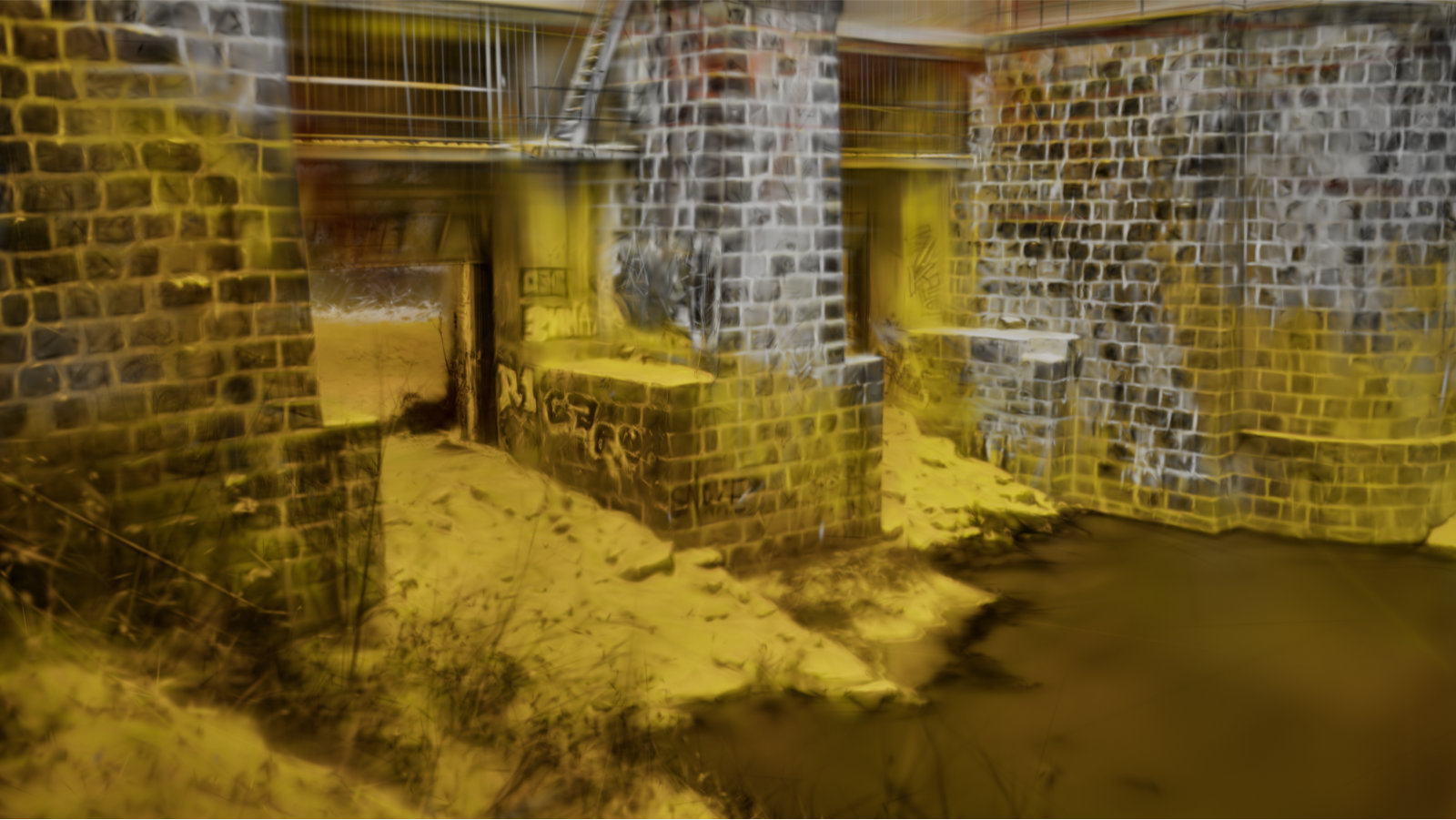}
        \caption{$[$\tOn[s3_3_1] \tOn[s3_3_2] \tOn[s3_3_3]$]$}
    \end{subfigure}
    \begin{subfigure}[b]{0.245\textwidth}
        \centering
        \includegraphics[width=\textwidth]{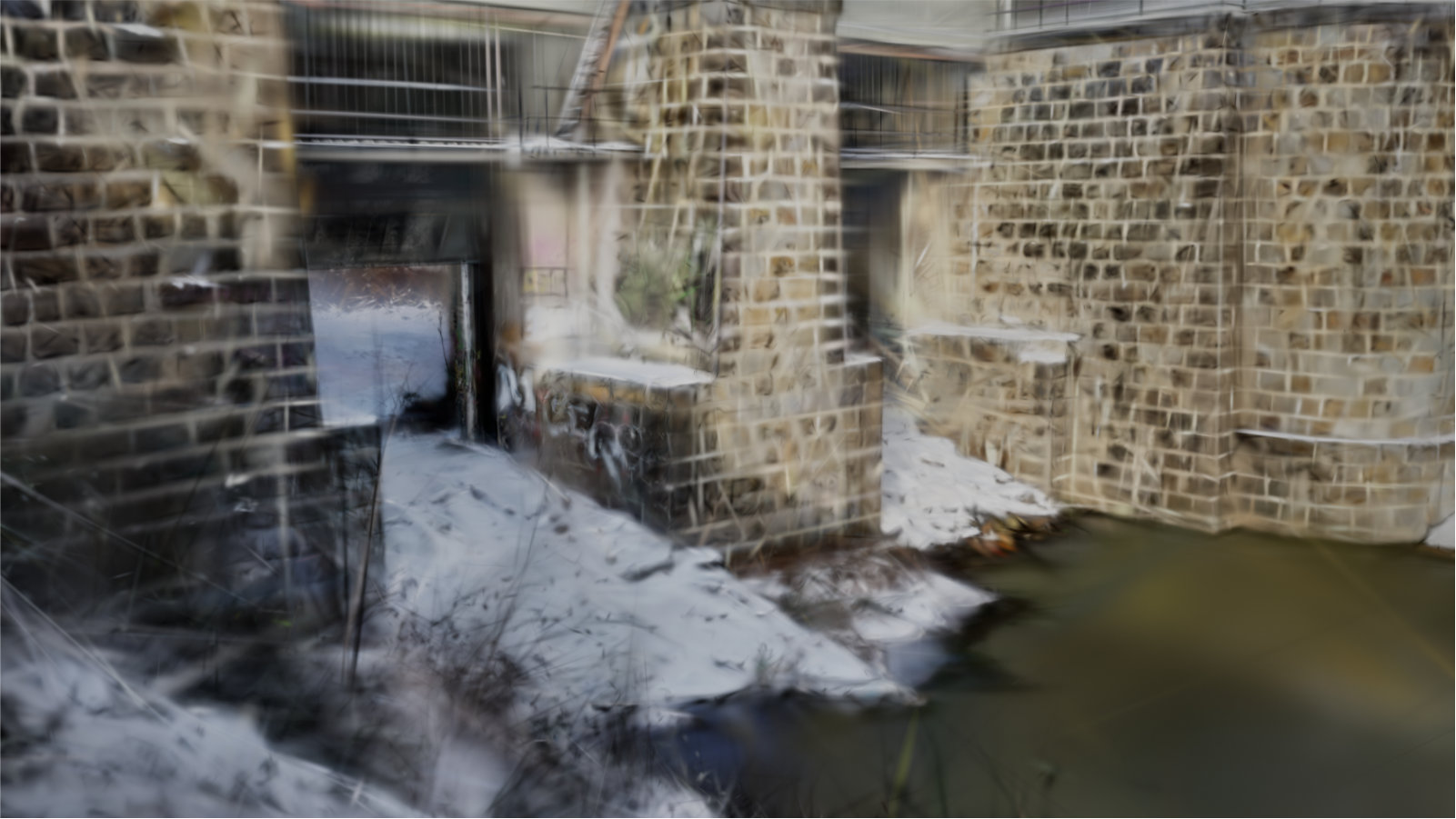}
        \caption{$[$\tOn \tOff \tOff$]$}
    \end{subfigure}
    \begin{subfigure}[b]{0.245\textwidth}
        \centering
        \includegraphics[width=\textwidth]{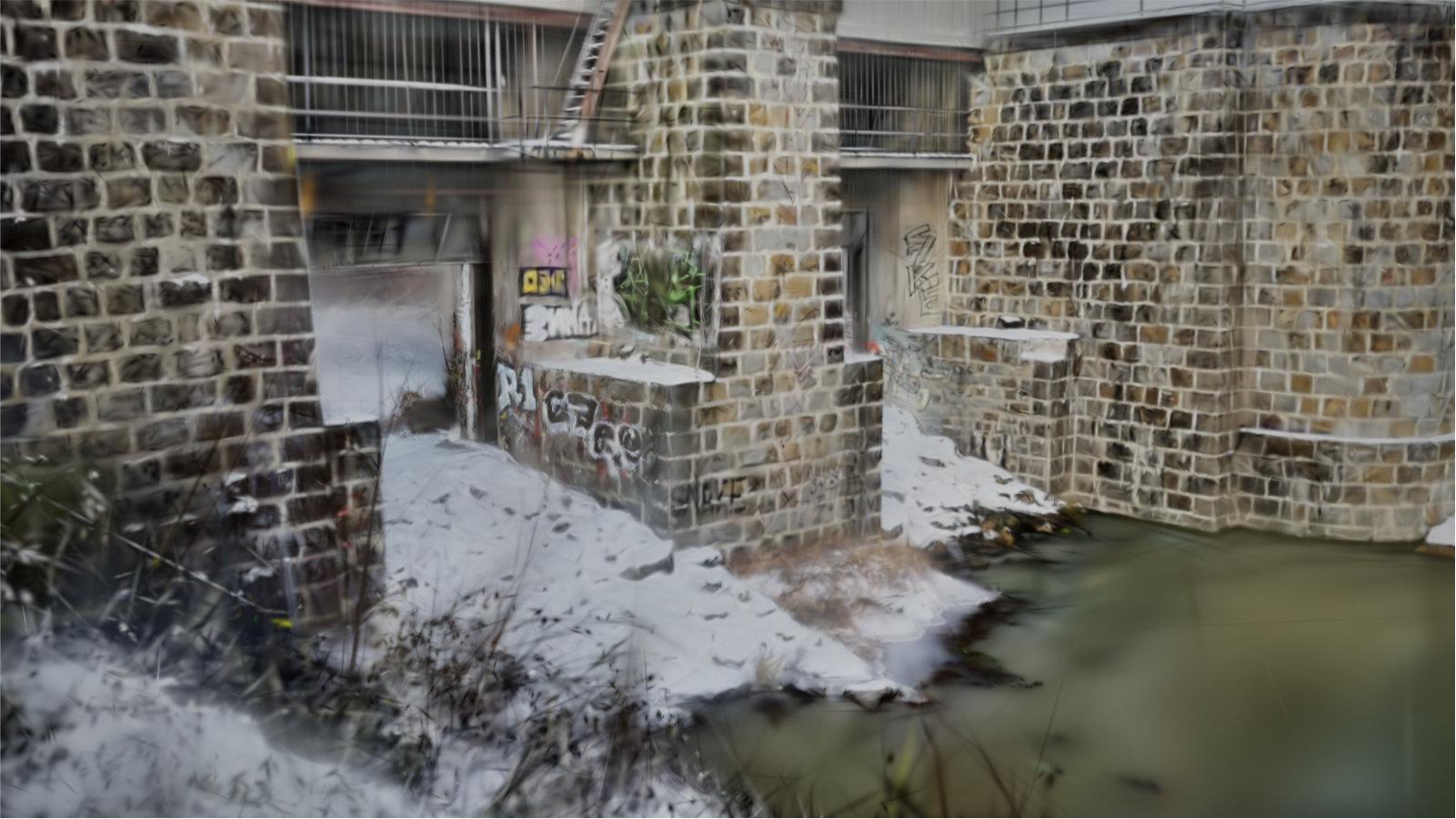}
        \caption{$[$\tOff \tOn \tOff$]$}
    \end{subfigure}
    \begin{subfigure}[b]{0.245\textwidth}
        \centering
        \includegraphics[width=\textwidth]{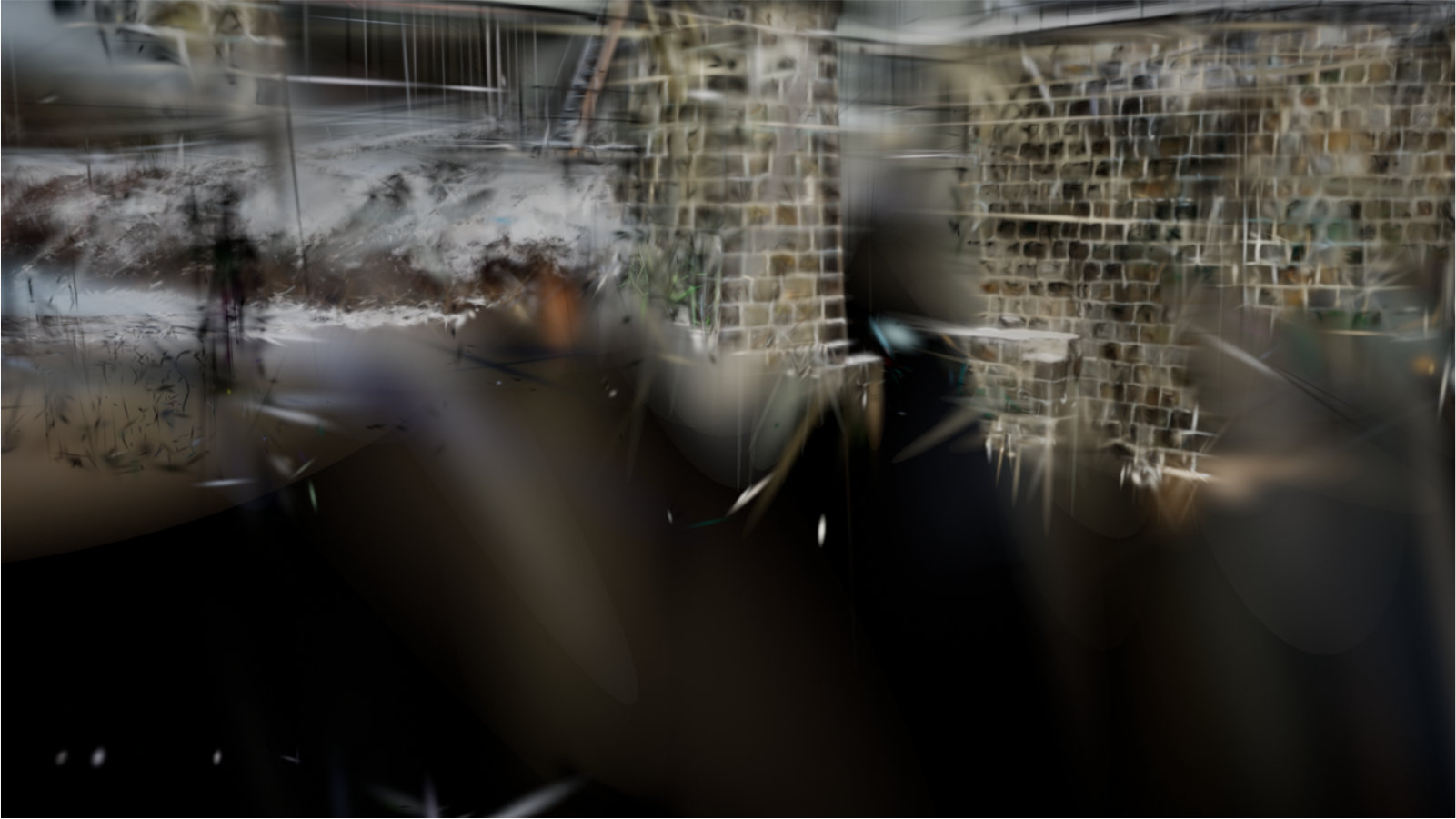}
        \caption{$[$ \tOff \tOff \tOn$]$}
    \end{subfigure}
    \caption{
Demonstration of our change visualization on regions not covered by the input data at one timestep. (a) Yellow highlights areas not covered at $t_3$ but present at $t_1$ and $t_2$; grayscale indicates regions persistent across all timesteps. (b)--(d) Individual timestep renderings for $t_1$--$t_3$.
 \tOn indicates an enabled timestep, with color corresponding to timestep presence in (a); \tOff indicates a disabled timestep.
}
    \label{fig:missing_area}
\end{figure*}

\section{Discussion}
Our results show that we are able to reliably fuse individual timesteps through our approach. The cross-timestep optimization leads to higher levels of detail compared to individually trained Gaussian sets without increasing the combined Gaussian count. By allowing new pre-trained Gaussian sets to be added incrementally, the reconstruction can be improved over time without reprocessing previously merged timesteps. Through the integrated persistence encoding at the Gaussian primitive level, our visualization approach highlights changes at sub-object granularity and indicates at which timesteps parts of the scene are present. 
Our results demonstrate reliable change highlighting across both natural environments~--~including vegetation, snow, and water levels~--~and built environments, even in sparsely covered areas (Figure~\ref{fig:highlighting_diff}).
By fusing different Gaussian sets, users can explore scenes without switching between individual reconstructions. Our approach provides visual indications of at which timestep a subarea can be validated and which areas have changed -- critical information for use cases such as mission planning and decision making in case of natural disasters. Additionally, the change visualization can support capture planning by highlighting which structures or parts of structures have not been captured in recent timesteps, as illustrated in Figure~\ref{fig:missing_area}.

\section{Limitations and Future Work}
While we account for changes between timesteps, we do not consider structural changes or lighting shifts that occur within a single recording period. 
Our approach also depends on a sufficiently accurate SfM registration when merging initial Gaussian sets. If the scene changes dramatically between timesteps (e.g., due to snow cover), the registration may fail, preventing any improvement in reconstruction quality.  \added{Additionally, the SfM registration is a time-consuming preprocessing step and more efficient 3DGS-based registration methods~\cite{chang2024gaussreg} could substantially reduce the time required to incorporate new timesteps.}
\replaced{Our approach currently preserves approximately the same number of Gaussians throughout the optimization, due to the lack of a cross-timestep pruning step. We believe that adding such a step could substantially reduce the Gaussian count without compromising reconstruction quality or persistence encoding, which we leave as future work.}
{ We have not implemented an efficient pruning step, and, as a result, our approach preserves the number of Gaussians throughout the optimization. For scenes with little change and strongly overlapping Gaussian sets, we believe the Gaussian count could be reduced substantially without compromising reconstruction quality or persistence encoding. We aim to address this in future work.}

Our user study evaluated the visualization approach on pre-rendered 2D images, isolating the perceptual task from 3D navigation and interface familiarity effects, which we consider a natural next step toward a realistic field deployment evaluation.

\section{Conclusion}
We presented ChronoFuseGS, a multi-temporal Gaussian Splatting approach that merges individually pre-trained Gaussian models into a single refined reconstruction supporting novel-view synthesis across multiple timesteps. By encoding per-Gaussian persistence, our approach leverages data across all captured timesteps, improving the reconstruction quality over individually trained models, while enabling incremental extension as new data becomes available. Our change-aware visualization highlights scene changes at sub-object granularity and indicates at which timesteps individual parts of the scene are present, without requiring users to switch between individual reconstructions. We demonstrated our approach on a real-world outdoor dataset spanning 7 months, showing consistent improvements in novel-view synthesis quality and effective change highlighting.

\section{Acknowledgements}
This research was funded in part by the Austrian KIRAS program of the Federal Ministry of Finance within the PostDisaster project.

\bibliographystyle{eg-alpha-doi} 
\bibliography{manuscript/gaussians}


\end{document}